\documentclass[showpacs,aps,prd,floatfix,amsmath,amssymb]{revtex4}

\usepackage{cleveref}
\usepackage{multirow}
\usepackage{amsmath}
\usepackage{amsfonts}
\usepackage{amssymb}
\usepackage{subfigure}
\usepackage[utf8]{inputenc}
\usepackage{graphicx}
\usepackage{color}
\usepackage{dcolumn}
\usepackage{bm}
\usepackage{float}

\catcode`\@=11
\def\lsim{\mathrel{\mathpalette\@versim<}}
\def\gsim{\mathrel{\mathpalette\@versim>}}
\def\@versim#1#2{\vcenter{\offinterlineskip
\ialign{$\m@th#1\hfil##\hfil$\crcr#2\crcr\sim\crcr } }}
\catcode`\@=12
\catcode`@=12

\begin{document}
\title{Searching for LFV Flavon  decays at hadron colliders }

\author{
M. A. Arroyo-Ure\~na}
\affiliation{
   Facultad de Ciencias F\'{\i}sico-Matem\'aticas, Benem\'erita Universidad Aut\'onoma de Puebla, Puebla, M\'exico.
}

\author{
A. Bola\~nos}
\affiliation{Departamento de Ciencias e Ingenierías, Universidad Iberoamericana, Campus Puebla, Puebla, M\'exico.
}

\author{J. L. D\'iaz-Cruz}
\affiliation{
   Facultad de Ciencias F\'{\i}sico-Matem\'aticas, Benem\'erita Universidad Aut\'onoma de Puebla, Puebla, M\'exico.
}

\author{
G. Hern\'andez-Tom\'e}
\affiliation{
Departamento de F\'{\i}sica, CINVESTAV
IPN, Apartado Postal 14-740, 07000, M\'exico D. F., M\'exico.
}

\author{
G. Tavares-Velasco, }
\affiliation{
   Facultad de Ciencias F\'{\i}sico-Matem\'aticas, Benem\'erita Universidad Aut\'onoma de Puebla, Puebla, M\'exico.
}

  \date{\today}

\begin{abstract}
The search for Flavons with a mass of $\mathcal{O}$(1) TeV at current and future colliders might probe low-scale
flavor models. We are interested in the simplest model that invokes the Froggatt-Nielsen (FN) mechanism with an Abelian
flavor symmetry, which includes a Higgs doublet  and a FN  complex singlet. Assuming a CP conserving scalar potential,
there are a $CP$-even  $H_F$ and a $CP$-odd $A_F$ Flavons with lepton flavor violating (LFV) couplings.  The former can mix with the standard-model-like Higgs boson, thereby  inducing tree-level LFV  Higgs interactions that may be at the reach of the  LHC.
We study the constraints on the parameter space of the model from low-energy LFV processes,
which are then used to evaluate the Flavon decay widths  and the $gg\to \phi \to \tau\mu$ ($\phi=H_F,\,A_F$) production cross section  at
hadron colliders. After imposing several kinematic cuts to reduce the SM main background, we find that for $m_{H_F}$ about 200-350 GeV, the decay $H_F \to \tau \mu$ might be at the reach of the LHC
for a luminosity in the range  1-3 ab$^{-1}$, however, a  luminosity of the order of 10 ab$^{-1}$ would be required
to detect the $A_F \to \tau \mu$ decay. On the other hand a future 100 TeV $pp$
collider could  probe masses as high as $\mathcal{O}$(10) TeV if it reaches an integrated luminosity
of at least 20 ab$^{-1}$. Therefore, the 100 TeV Collider could work as a Flavon factory.
\end{abstract}

\maketitle


\section{Introduction}

After the discovery  of a Higgs-like particle with a mass $m_h=125-126$ GeV
\cite{Aad:2012tfa,Chatrchyan:2012xdj}, the search for new physics (NP) has become the
one of the next goals of the LHC.  Although current measurements of the spin, parity, and couplings of the Higgs boson
seem  consistent with the standard model (SM) \cite{Gunion:1989we}, its light mass seems troublesome, i.e. the hierarchy problem, and calls for new physics (NP). The SM has also other
open issues, such as the flavor problem, unification, etc. \cite{Pomarol:2012sb,Martin:1997ns}, which
also encourages the study of NP models.

The couplings of the Higgs particle to a pair of massive gauge bosons or fermions
have  strengths proportional  to the masses of such particles.  However,  the LHC has tested only a few of such Higgs couplings,
namely the ones to the gauge bosons and the  heaviest  fermions. Along these lines,
many studies have been devoted to analyze the pattern of Higgs couplings derived from LHC data,
for instance \cite{Espinosa:2012ir,Giardino:2013bma}. However, non-standard Higgs couplings,
including the flavor violating (FV) ones, are  predicted in many models of physics beyond the SM
\cite{Branco:2011iw, DiazCruz:2004tr, DiazCruz:2004pj, DiazCruz:2002er}. In particular, the observation of neutrino oscillations,
which is associated with massive neutrinos, motivates the occurrence of lepton flavor violation (LFV)
in nature \cite{Ma:2009dk}.
Within the SM, LFV processes vanish  at any order of perturbation theory, which  motivates
the study of  SM extensions that predict sizeable  LFV effects that could be at the reach of detection.
Apart from  decays such as $l_i\to l_j \gamma$ and $l_i\to l_j \bar l_k l_k$, particularly interesting
is the  decay $h \to \tau \mu$, which was studied first in Refs. \cite{Pilaftsis:1992st, DiazCruz:1999xe},
with subsequent analyses on the detectability of the signal appearing soon after
\cite{Han:2000jz, Assamagan:2002kf}. This motivated a plethora of calculations
in the framework of several SM extensions, such as theories with massive neutrinos, supersymmetric
theories, etc.  \cite{Arganda:2004bz,DiazCruz:2002er,Brignole:2004ah,DiazCruz:2008ry,Chamorro-Solano:2017toq,Chamorro-Solano:2016ugt,Lami:2016mjf}.
After the Higgs boson discovery, the decay $h\to\tau\mu$  offers a great opportunity to search for
NP at the LHC. Although a slight excess of  the $h \to \tau \mu$ branching ratio was reported at the
LHC run I, with a significance of 2.4 standard deviations \cite{Khachatryan:2015kon},
 a  subsequent study \cite{CMS:2016qvi} ruled out such an  excess and put the limit
$BR(h\to \bar{\mu}\tau) < 1.2 \times 10^{-2}$  with 95\% C.L. In the model we are interested in,
LFV effects are induced at the tree-level in the scalar sector, so it is thus worth assessing their phenomenology.

Another open  issue in the SM is the flavor problem \cite{Isidori:2010kg}, which  has long been the focus of interest, with
several proposals meant to address it, such as
textures, GUT-inspired relations,  symmetries, radiative generation, etc.
In particular, a flavor symmetry approach can be supplemented with the Froggatt-Nielsen (FN) mechanism, which
assumes that above some scale $M_F$,  there is a symmetry (perhaps of Abelian type $U(1)_F$) that forbids the  appearance of Yukawa couplings; SM fermions are charged under this symmetry.
However, the Yukawa matrices can arise  through  non-renormalizable operators.
The Higgs spectrum of these models could include a light   Flavon  $H_F$, which could mix with
the Higgs bosons when the flavor scale is of the order of the TeVs.
Quite recently,  the phenomenology  of Higgs-Flavons at particle colliders has  been the focus of attention
 \cite{Bolanos:2016aik, Bauer:2016rxs, Huitu:2016pwk, Berger:2014gga, Diaz-Cruz:2014pla}.

Although the states we are interested in arise from the mixing of
Higgs bosons and Flavons, we will still call them Flavons for short.  Depending on the particular model, there could be
several potentially detectable Flavon  decays, which would be indistinguishable from the decays of a heavy Higgs boson, therefore, to search for a distinctive signature, we will focus
on the one arising  from the LFV decay   $H_F\to \tau\mu$,  with $\tau\mu=\tau^- \mu^+ +\tau^+\mu^-$.
The minimal model that  introduces the FN mechanism with an Abelian flavor symmetry
includes a scalar sector consisting of a Higgs doublet and a FN  complex singlet. From now on we will
refer to this model as the FN extension of the Standard Model  (FNSM). Such a model predicts a $CP$-even Flavon $H_F$
and a $CP$-odd one $A_F$. Also, the couplings of the light SM-like Higgs boson would deviate from the SM ones, such that  two possible scenarios are possible: firstly, the mixing of the real part of the doublet with the real
component of the FN singlet could induce sizable LFV Higgs couplings of the light physical
Higgs boson, which might affect the light Higgs phenomenology; secondly, the $CP$-even Flavon
could be very heavy, so its mixing with the light Higgs boson would be negligible and unconstrained
by LHC Higgs data, though in such a case the $CP$-odd state $A_F$ could be the lighter one, thereby  giving rise to a potentially
detectable LFV signal.

In this paper  we are interested in studying the possible detection of both $CP$-even and $CP$-odd Flavons  at the LHC  and a future
100 TeV $pp$ collider via their LFV decays.  It has been pointed out that a 100 TeV $pp$ collider would allow for a detailed study of several topics of interest in particle physics, such as  Higgs physics and the electroweak symmetry breaking mechanism    \cite{Contino:2016spe}. It will also be useful to search for possible signals of dark matter, SUSY theories, and other extension models \cite{Golling:2016gvc}.

We will start our analysis by considering the constraints on the parameter space of the FNSM obtained from Higgs data at the LHC,  low energy LFV, and the muon magnetic dipole moment. A set of benchmarks will then be used to estimate the Flavon decay modes, focusing on the LFV ones, as well as their production cross sections by gluon fusion at the LHC and a future 100 TeV $pp$ collider.  We will then explore the possibility  that the $CP$-even and $CP$-odd  Flavons  could be detected via the $\tau\mu$ decay channel, for which we will make a Monte Carlo analysis of the signal and the SM main background.

The organization of our work is as follows. In Sec. II we describe the realization of the FN mechanism
within the simplest model, namely that  with  one Higgs doublet and one FN singlet.  In particular,
we present the Higgs potential and Yukawa Lagrangian, from which the Flavon  couplings can be extracted.
Section III is devoted to the constraints on the parameters space of the model and the benchmarks of parameter values we will be using in our analysis, whereas   Sec. IV is focused on the analysis of the decay modes of both a $CP$-even and a $CP$-odd Flavon as well as their production cross sections via gluon fusion at the
LHC and a future 100 TeV $pp$ collider.  We  also present the Monte Carlo analysis of the $\mu\tau$ signal and its main background. The conclusions and outlook are presented in Sec. V.


\section{The scalar sector of the minimal FNSM}

The scalar sector of the FNSM includes the usual SM Higgs doublet

\begin{equation}
 \Phi = \left( \begin{array}{c} G^+ \\ \frac{1}{\sqrt{2}} \left( v + \phi^0 + i G_Z \right)\\
\end{array} \right),
 \label{dec_doublets}
\end{equation}
and a complex singlet
\begin{equation}
S_F = \frac{1}{\sqrt{2}} (u + s + i p ),\label{dec_singlet}
\end{equation}
where $v$ denotes the SM vacuum  expectation value (VEV) and $u$  that  of the FN singlet, whereas
$G^+$ and $G^0$ will be identified with the pseudo-goldstone bosons that  become the longitudinal modes of
the $W$ and $Z$ gauge bosons.

\subsection{The Higgs potential}

We turn now to discuss the minimal $CP$-conserving Higgs potential with a softly-broken $U(1)$ global symmetry, which is given as follows

\begin{align}
V &= -\frac{1}{2} {m_{1}^2} \Phi^\dagger \Phi -\frac{m_{s_1}^2}{2} S^*_F S_F -\frac{m_{s_2}^2}{2} (S^{*2}_F + S^2_F)\nonumber\\
&+ \frac{1}{2} \lambda_1 \left( \Phi^\dagger \Phi \right)^2
+ \lambda_{s} (S^*_F S_F)^2
+ \lambda_{11} (\Phi^\dagger\Phi) (S^*_F S_F).
\label{potIDM1S}
\end{align}
We are therefore left with the following $U(1)$-symmetric terms
($m_{1}^2, m_{s_1}^2, \lambda_{1}, \lambda_{s}, \lambda_{11}$), and the $U(1)$-soft-breaking
term $m_{s_2}^2$. The latter is required to avoid a massless Goldstone boson when $\langle S_F\rangle\ne 0$. An extensive analysis of this potential was presented in reference \cite{Bonilla:2014xba},
where the parameter space that allows a viable model was identified.

After imposing the minimization conditions on the potential,  the following relations are obtained:

\begin{eqnarray}
 m_{1}^2  &=&  v^2 \lambda_1 + u^2 \lambda_{11}, \\
 m_{s_1}^2 &=&  -2m_{s_2}^2+2 u^2 \lambda_{s} + v^2 \lambda_{11}.
 \end{eqnarray}


\subsection{The Scalar Mass Matrix}

In a $CP$-invariant potential, the $CP$-even (real)  and $CP$-odd (imaginary)
components of the mass matrix do not mix.
In this case the mass matrix for the real components in the basis $(\phi^0,s)$ is given by:

\begin{equation}
M^2_S =
 \left( \begin{array}{cc}
   \lambda_1 v^2      &  \lambda_{11} u v \\
\lambda_{11} u v    &   2 \lambda_{s} u^2
\end{array} \right),
\end{equation}
whereas the mass matrix for the imaginary components, in the basis $(G_Z,p)$, reads

\begin{equation}
M^2_P =
 \left( \begin{array}{cc}
   0       &  0     \\
   0      & 2 m^2_{s_2}
\end{array} \right).
\end{equation}
We notice that the mass scale for the $CP$-odd state arising from the FN singlet $A_F =p$
is different from the VEV $u$, which is the $U(1)$-breaking scale, and therefore it could be much lighter.
As for the mixing of the real components of the doublet $\Phi$ and the singlet $s$,
the mass eigenstates are obtained through the standard $2\times 2$ rotation:

\begin{eqnarray}
 \phi^0   &=& \, \, \, \, \cos \, \alpha \, h + \sin \, \alpha  \, H_F, \label{phi0mix} \\
  s    &=& -\sin \, \alpha \, h + \cos \, \alpha \, H_F. \label{smix}
\end{eqnarray}

In what follows we will identify the mass eigenstate $h$ as the SM-like
Higgs boson with $m_h=125$ GeV, while the mass eigenstates  $H_F$ and $A_F$ will be assumed to be heavier.
Although they arise from Flavon-Higgs mixing, in the present work we will still refer to  $H_F$ and $A_F$ as Flavons for short.
The properties of the $CP$-even Flavon   will depend on the  size of its mixing with the lightest state. On the other hand,
the $CP$-odd state, which   does not couple to gauge bosons, it does couple to the SM fermions,
including both diagonal and non-diagonal interactions.

Our analysis of Flavon  decays requires the knowledge of  cubic interactions,
such as the trilinear vertex $H_F hh$, which is given in the minimal model by:

\begin{equation}
 g_{H_F hh}  =  \frac{1}{2} [  \lambda_{11}  (u \cos^3 \alpha  + v \sin^3 \alpha )
                       + 2 u \sin^2 \alpha \cos \alpha (3\lambda_{s} -\lambda_{11})
                       + v \sin \alpha \cos^2 \alpha (3\lambda_1 - 2\lambda_{11}) ].
\end{equation}

\subsection{Yukawa sector and LFV interactions}

The FN Lagrangian of the model includes the terms that   become the Yukawa
couplings once the $U(1)_F$ flavor symmetry is spontaneously broken. It is given by:
\begin{equation}
 -{\cal{L}}_Y =    \rho^d_{ij} \left( \frac{ S_F }{\Lambda_F} \right)^{q^d_{ij}} \bar{Q}_i d_j  \tilde{\Phi}
                + \rho^u_{ij}  \left(\frac{ S_F }{\Lambda_F}\right)^{q^u_{ij}} \bar{Q}_i u_j \Phi
                + \rho^l_{ij}  \left(\frac{ S_F }{\Lambda_F}\right)^{q^l_{ij}} \bar{L}_i l_j \Phi  +{\rm H.c.},
\end{equation}
where $q^f_{ij}$ $(f=u,d,l)$ denote the   Abelian charges that reproduce the observed
fermion masses, for each fermion type.
The Flavon  field $S_F$ is assumed to have flavor charge equal to -1, such that
${\cal{L}}_{Y}$ is $U(1)_F$-invariant.
Then,  the Yukawa couplings arise after the spontaneous breaking  of the flavor symmetry,
i.e. $\lambda_x = (\frac{<S_F>}{\Lambda_F})^{n_x}$, where $<S_F>$ denotes the
Flavon  VEV, whereas $\Lambda_F$ denotes a heavy mass
scale, which represents the mass of heavy fields that
transmit such symmetry breaking to the quarks and leptons.
For a detailed discussion of the viable structures for the Yukawa matrices for the FN multi-Higgs model see Ref. \cite{Bauer:2016rxs}. Here we shall discuss the generic features that can be identified by studying specific Abelian charges for the charged leptons.
With this purpose we consider the charge assignment used by one of us in Ref. \cite{Diaz-Cruz:2014pla}, where the Yukawa matrix $Y^l$ is of the form:
\begin{equation}
Y^{l}
\sim\begin{pmatrix}
\lambda^{6} & \lambda^{6} & \lambda^{6}\\
\lambda^{6} & \lambda^{4} & \lambda^{4}\\
\lambda^{6} & \lambda^{4} & \lambda^{2}
\end{pmatrix},
\end{equation}
which can be justified with the following  Abelian charges
\begin{itemize}
\item $\alpha_i$= $U(1)_F$ charge of lepton doublet $L_i$,
\item $\beta_j$= $U(1)_F$ charge of lepton singlet $l_j$.
\end{itemize}
Then, $Y_{ij}^{l}\sim\lambda^{|\alpha_i-\beta_j|}$, which means that choosing $Y_{23}^{l}\sim\lambda^4$ and $Y_{33}^{l}\sim\lambda^2$  implies $|\alpha_2-\beta_3|=4$ and $|\alpha_3-\beta_3|=2$. 
Other examples of Abelian charges are presented in Ref. \cite{Bauer:2016rxs,Baldes:2016gaf,Grossman:1995hk}. We then focus on the 2-3 submatrix, which dictates the $\phi\tau\mu$ interaction. The corresponding squared mass matrix can be written as:
\begin{equation}
M_{l}^{2}=v^{2}|Y_{33}^{l}|^{2}
\begin{pmatrix}
\lambda^{2} & \lambda^{2}\\
\lambda^{2} & 1
\end{pmatrix},
\end{equation}
with $Y_{33}^{l}\sim \lambda^2=(m_{\tau}-m_{\mu})/v$. This mass matrix can be diagonalized by a $2\times 2$ rotation with mixing angle $\theta\sim(m_{\mu}/m_{\tau})$.

In  the unitary gauge we set   $G^{\pm}\rightarrow 0$, $G_Z \rightarrow 0$, thus
we can write   the
doublet $\Phi$ as follows:
	\begin{equation}
		\Phi =\frac{1 }{\sqrt{2}} \begin{pmatrix}
		           0\\
v + \phi^0 \end{pmatrix},
	\end{equation}
whereas the powers of the Flavon  field can be expanded as
	\begin{eqnarray}
		S^{q^f_{ij}} = \left( \frac{u+ s + i p}{\sqrt{2}} \right)^{q^f_{ij}} \simeq
		\left( \frac{u}{\sqrt{2}} \right)^{q^f_{ij}} \left[ 1 +q^f_{ij}\left(\frac{s+i p}{u} \right) \right].
	\end{eqnarray}
Then, after substituting the Higgs boson and Flavon  mass eigenstates,  one gets finally the following interaction Lagrangian for
the Higgs-fermion couplings

\begin{eqnarray}
 -{\cal{L}}_Y &=&   \frac{1 } {  v} [ \bar{U} \tilde M_u U  +  \bar{D}  \tilde M_d  D
             +  \bar{L} \tilde M_l L ] ( c_{\alpha} h + s_{\alpha} H_F) \nonumber \\
            & & +  \frac{v } { \sqrt{2} u}  [\bar{U}_i \tilde{Z}^u_{ij} U_j  +  \bar{D}_i \tilde{Z}^d_{ij} D_j
                +  \bar{L}_i\tilde{Z}^l_{ij}  L_j ]  ( -s_{\alpha} h + c_{\alpha} H_F + i A_F) +{\rm  H.c.},
\end{eqnarray}
where we use the usual short-hand notation $ s_{\alpha}\equiv\sin\alpha$ and $ c_{\alpha}\equiv\cos\alpha$. Here, $\tilde M^f$ is the diagonal mass matrix, whereas the information about the size of FV Higgs-Flavon couplings is contained in the $\tilde{Z}^f=U_L^f{Z}^fU_L^{f\dagger}$ matrices, with  $Z^f_{ij}$ given in the flavor basis as
\begin{equation}
\label{Zij}
Z^f_{ij}=\rho^f_{ij}\left(\frac{u}{\sqrt{2}\Lambda_F}\right)^{q^f_{ij}}q^f_{ij},
\end{equation}
which remains non-diagonal once the fermion mass matrices are diagonalized, thereby giving rise to FV scalar couplings.
For the Yukawa matrix we are considering, the $Z$ matrix is given as follows (for the 2-3 subsystem):
\begin{equation}
Z^{l}=\begin{pmatrix}
4Y_{22}^{l} & 4Y_{23}^{l}\\
4Y_{32}^{l} & 2Y_{33}^{l}
\end{pmatrix}.
\end{equation}
Thus we can use the approximate diagonalization of the mass matrix to express $Y_{ij}^{l}$ and $Z^{l}$ in terms of the mass ratios. To leading order we have 
\begin{equation}
\widetilde{Z}^{l}\backsimeq\begin{pmatrix}
4\frac{m_{\mu}}{v}\frac{\sqrt{m_{\mu}}}{m_{\tau}} & 6\frac{m_{\mu}}{v}\\
6\frac{m_{\mu}}{v} & 2\frac{m_{\tau}}{v}
\end{pmatrix},
\end{equation}
and similary for  up-type quarks, whose Higgs couplings play a fundamental role for collider phenomenology. With the assignment of Abelian charges chosen in \cite{Diaz-Cruz:2014pla}, the $Z^u$ matrix for the 2-family case is given by:
\begin{equation}
Z^u=
\begin{pmatrix}
2Y^u_{22}&2Y^u_{23}\\
2Y^u_{32}&0
\end{pmatrix},
\end{equation}
with $Y^u_{22}\sim Y^u_{23}\sim \lambda^2$. However, in the mass-eigenstates basis, the rotated matrix takes the following form to leading order:
\begin{equation}
\widetilde{Z}^u \sim
\begin{pmatrix}
2Y^u_{22}&2Y^u_{23}\\
2Y^u_{32}&4 s_u Y^u_{23}
\end{pmatrix},
\end{equation}
where $s_u=\sin \theta_u$, with $\theta_u$ the mixing angle. It  turns out that the $\phi t\bar{t}$ coupling vanishes in the weak basis, but it is non-vanishing as long as there is mixing between the top quark and the mass eigenstates basis.
The diagonal and non-diagonal interactions of the $h$, $H_F$, and $A_F$ scalar bosons with massive fermions
are given by:
\begin{align} \label{ghff}
g_{hf_i f_j} &= \frac{c_{\alpha} }{v} \tilde M^f_{ij} - s_{\alpha} r_s {\tilde{Z}^f}_{ij},\nonumber \\
g_{H_F f_i f_j } &= \frac{s_\alpha }{v} \tilde M^f_{ij} + c_\alpha r_s {\tilde{Z}^f}_{ij},  \nonumber \\
g_{A_F f_i f_j} & =  i r_s {\tilde{Z}^f}_{ij},
\end{align}
where the Feynman rule for the $A_F f_i f_j$ vertex includes a $\gamma^5$ Dirac matrix and $r_s= v/(\sqrt{2} u)$.

Besides the Yukawa couplings, we  also need to specify the scalar-to-gauge-boson couplings. They
can be readily extracted from the kinetic terms of the Higgs doublet and the singlet, which
transforms trivially under the SM gauge group. Thus after substituting Eq. \eqref{phi0mix} in the kinetic term, we obtain
that the $h$ and $H_F$ couplings to gauge boson pairs are SM-like, with the coupling constants given
by $g_{h_i VV}=\chi_V^{h_i} g_{h_{SM} VV}$ $(V=W,Z)$, for $h_i=h,\,H_F$,
with $\chi^h_V\;(\chi^{H_F}_V)=\cos\alpha\;(\sin\alpha)$.
Thus the coupling constants are

\begin{align} \label{ghVV}
g_{hZZ}&=\frac{g m_Z}{c_W} \cos\alpha\\
g_{hWW}&=g m_W \cos\alpha,\\
g_{H_FZZ}&=\frac{g m_Z}{c_W} \sin\alpha,\\
g_{H_F WW}&=g m_W \sin\alpha.
\end{align}

We are interested in the possible detection of $CP$-even and $CP$-odd Flavons with masses of the TeV order at both
the LHC and a future $pp$ collider with a center-of-mass energy of $100$ TeV.
Depending on the particular model, there could be several potentially detectable decays of such Flavons,
but some of them would also arise from  heavy Higgs bosons, for instance
within multi-Higgs doublet models. Thus, in order  to search for a distinctive Flavon  signature, we shall focus
on the one arising  from the LFV decay   $\phi \to  \tau \mu$ ($\phi=H_F,\, A_F$).
In order to determine the detectability of this decay, we will present a Monte Carlo analysis of the signal and
the most relevant SM backgrounds.

\section{Constraints on the parameter space of the FNSM}

In order to  evaluate the  Flavon  decays and production modes at a hadron collider we need to analyze
the most up-to-date constraints on  the model parameters. For the mixing angle $\alpha$ we can use the data obtained
by the LHC collaborations  on the Higgs boson properties, whereas the LFV couplings can be constrained
via the experimental data on the muon anomalous magnetic dipole moment (AMDM) $a_\mu$, the LFV decays of the
tau lepton $\tau\to \bar{l_i}\bar{l_j}l_j$ and $\tau \to l_i \gamma$, as well as the experimental
constraint on the $h\to\mu\tau$  decay. All the necessary formulas to perform our analysis below are presented in  Appendix \ref{ConstraintFormulas}.

\subsection{Mixing angle $\alpha$}

 We shall use the universal Higgs
fit of Ref. \cite{Giardino:2013bma}, which presents constraints on the parameters $\epsilon_X$,
defined as  (small) deviations of the Higgs couplings from the SM values, i.e. $\chi^{h}_X=g_{hXX}/g_{hXX}^{SM}=1 + \epsilon_X$.
For the $W$ and $Z$ gauge bosons,  the corresponding constraints are $\epsilon_W= -0.15\pm 0.14$ and
$\epsilon_Z= -0.01\pm 0.13$. Regarding the fermion couplings, we notice that this universal Higgs fit is valid for
the $CP$-conserving case, which we are considering here; therefore we can apply them to constrain the properties of
the $CP$-even SM-like Higgs boson. Furthermore, the constraints derived from the gauge interactions provide
the strongest constraints on the mixing angle $\alpha$.
Since  the lightest  scalar boson $h$ couples with the SM gauge bosons with a strength that deviates from
the SM couplings by the factor   $c_{\alpha}$, to satisfy the bound on $\epsilon_Z$
we need to have  $0.86 < c_{\alpha} < 1$. We will use a conservative approach and  use the benchmark
 $c_\alpha=0.95$ in our analysis below.

\subsection{Diagonal $\tilde Z_{22}^{l}$ and $\tilde Z_{33}^l$ matrix elements}

\begin{figure}[htb!]
\centering
\addtocounter{subfigure}{0}
\includegraphics[width=13cm]{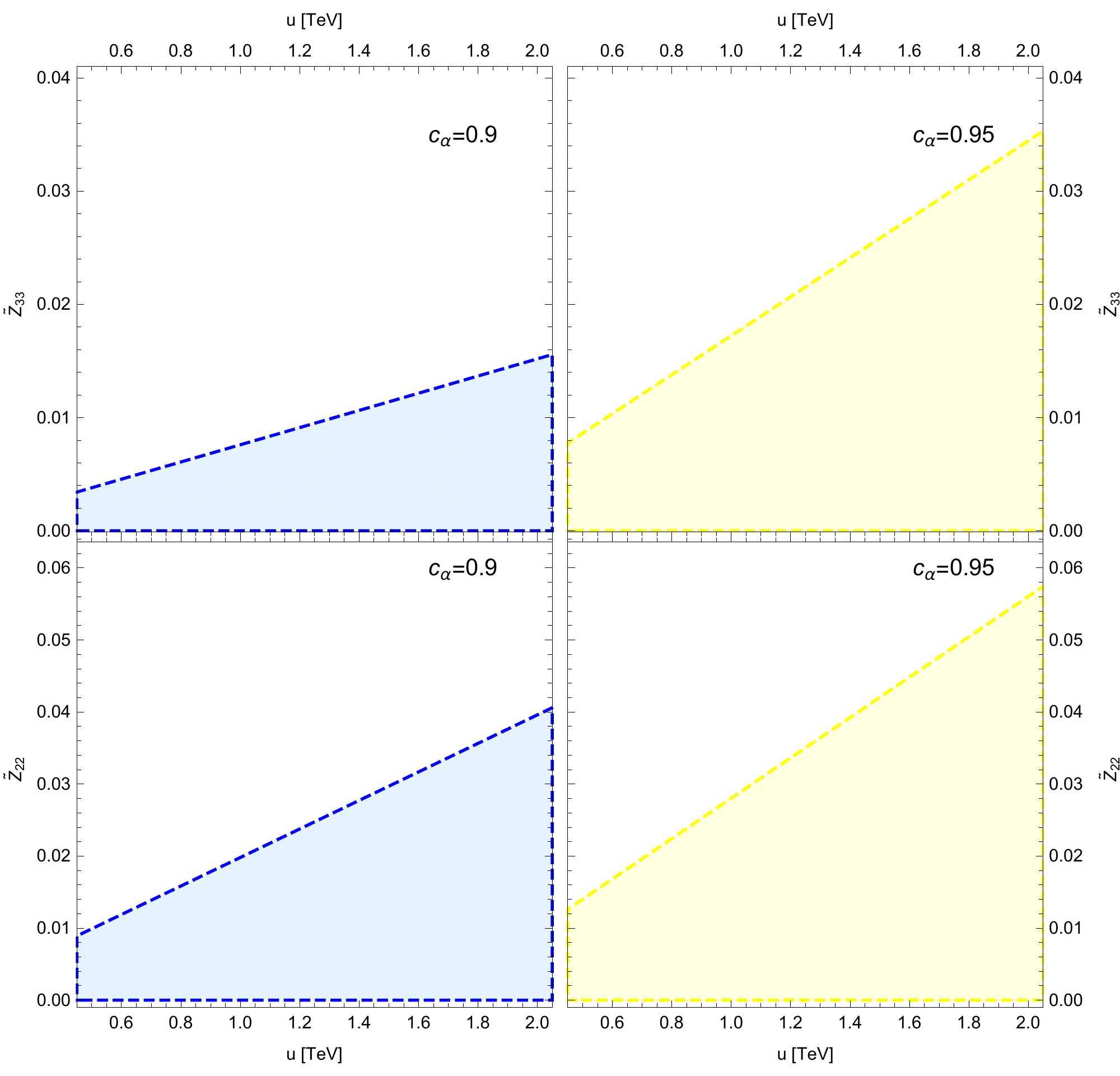}
\caption{Allowed values of  $\tilde Z_{33}^{l}$ (upper plots)  and  $\tilde Z_{22}^{l}$ (lower plots) as functions of $u$ and for two values of $c_\alpha$. To constraint $\tilde Z_{33}^{l}$ we use  the universal Higgs fit  on $\epsilon_\tau$ \cite{Giardino:2013bma}, whereas for $\tilde Z_{22}^{l}$ we use the ATLAS and CMS   bound $BR(h\to \bar\mu\mu)\lesssim 1.5 \times 10^{-3}$ with 95\% C. L.   \cite{Khachatryan:2014aep,Aad:2014xva,Khachatryan:2016vau}.}\label{Z33Constraint}
	\end{figure}

In this work we will use the 2-family approximation, neglecting FV with the leptons of the first fermion family.
We will also assume that there is no $CP$-violating phase, which means that we have three free parameters:
$\tilde Z_{22}$,  $\tilde Z_{23}$, and  $\tilde Z_{33}$.  To constrain the diagonal  $\tilde Z_{33}$
matrix element, we refer again to the  universal Higgs fit of Ref. \cite{Giardino:2013bma} and consider
the constraint on the  deviation of the SM $h \bar \tau \tau$ coupling, namely, $\epsilon_\tau= 0\pm 0.18$.
We thus show in Fig.    \ref{Z33Constraint} the allowed area on the $u-\tilde Z_{33}^l$ plane for
two values of $c_\alpha$. We observe that in order to agree with the universal Higgs fit constraint
when $c_\alpha=0.9$, $\tilde Z_{33}^l$ must be of the order of $10^{-3}$ for $u=0.5$ TeV and $10^{-2}$
for $u=2$ TeV, but when $c_\alpha=0.95$ we must have values of the order $10^{-2}$
in the complete $u$ interval. Since we are considering  $c_\alpha=0.95$, we will  use  $\tilde Z_{33}^l=10^{-2}$ as benchmark.

As far as $\tilde Z_{22}^l$ is concerned, the ATLAS and CMS Collaborations searched for the Higgs boson decay to $h\to \bar{\mu}\mu$ using the LHC data collected at $\sqrt{s} =7$ TeV and $\sqrt{s} =8$ TeV \cite{Khachatryan:2014aep,Aad:2014xva,Khachatryan:2016vau}. The upper bound  $BR(h\to \bar\mu\mu)\lesssim 1.5 \times 10^{-3}$ with 95\% C. L. was found for $m_h=125.6$ GeV, which is about one order of magnitude above the SM value $BR(h\to \bar\mu\mu)\simeq 2.04 \times 10^{-4}$. This bound would allow $\tilde Z_{22}^l$ values as large as $10^{-2}$ as shown in Fig. \ref{Z33Constraint}.  We will take instead a conservative approach and assume the hierarchy $\tilde Z_{22}^l< \tilde Z_{33}^l$, so we fix $\tilde Z_{22}^l=10^{-4}$, otherwise the SM $h \bar\mu\mu$ coupling would be swamped by the new corrections of the FNSM.

\subsection{Non-diagonal $\tilde Z_{23}^{l}$ matrix element}

We will consider   the current experimental bounds on  the muon AMDM $a_\mu$  \cite{Olive:2016xmw}, the tau decay $\tau\to\mu\gamma$ \cite{Olive:2016xmw},  and the Higgs boson decay $h\to\tau\mu$ \cite{Aad:2015gha,Khachatryan:2015kon} to constrain the   $Z_{ij}^{l}$ matrix elements \cite{Arroyo:2013tna}. Notice that  the current bound on the $\tau\to 3\mu$ decay width gives very weak constraints, so we  will omit such a process in our analysis.
Two scenarios arise when dealing with constraints on LFV couplings:

\begin{itemize}
  \item Scenario I: the FNSM is assumed to be responsible for the current discrepancy between the theoretical and experimental values of $a_\mu$, which requires a positive contribution from new physics. Along this line, the one-loop contribution from a $CP$-even scalar boson  is positive, whereas that of the $CP$-odd scalar boson is negative. Therefore a suitable scenario would be that with a relatively light $CP$-even Flavon and a heavy $CP$-odd Flavon, which would suppress the negative contribution. When $m_{H_F}\simeq m_{A_F}$, the Flavon  contributions would largely  cancel  one another out for $c_\alpha\simeq 1$ and the  remaining (positive)  contribution  would arise from the corrections to the SM Higgs boson couplings, unless there were large extra positive contributions arising from the $CP$ odd Flavon at the two-loop level: such contributions could arise as long as $A_F$ is very light and there was an enhanced one-loop-induced $A_F\gamma\gamma$ coupling, as discussed in \cite{Marciano:2016yhf}.  We do not expect such an enhancement in the FNSM, so large values of the mixing matrix element $Z_{23}^l$ would be required to solve the $a_\mu$ discrepancy.

  \item Scenario II: the FNSM Flavons give a negligible contribution to $a_\mu$ and fail to explain the discrepancy, though the model still can remain viable.   It could happen for instance  that  extra positive contributions to $a_\mu$ would arise from the ultraviolet completion of the FNSM, thereby solving the  puzzle. Also, it is still possible that  more precise determinations of the SM hadronic contribution and the experimental measurement would settle  the discrepancy in the future without requiring any NP effects.
       In any case, by requiring that the  Flavon contribution to $ a_\mu$ is negligible,  it is enough to satisfy the LHC constrain from $h\to\tau \mu$ in order to have a viable parameter space in this scenario.

\end{itemize}

\begin{figure}[htb!]
\centering
\addtocounter{subfigure}{0}
\includegraphics[width=15cm]{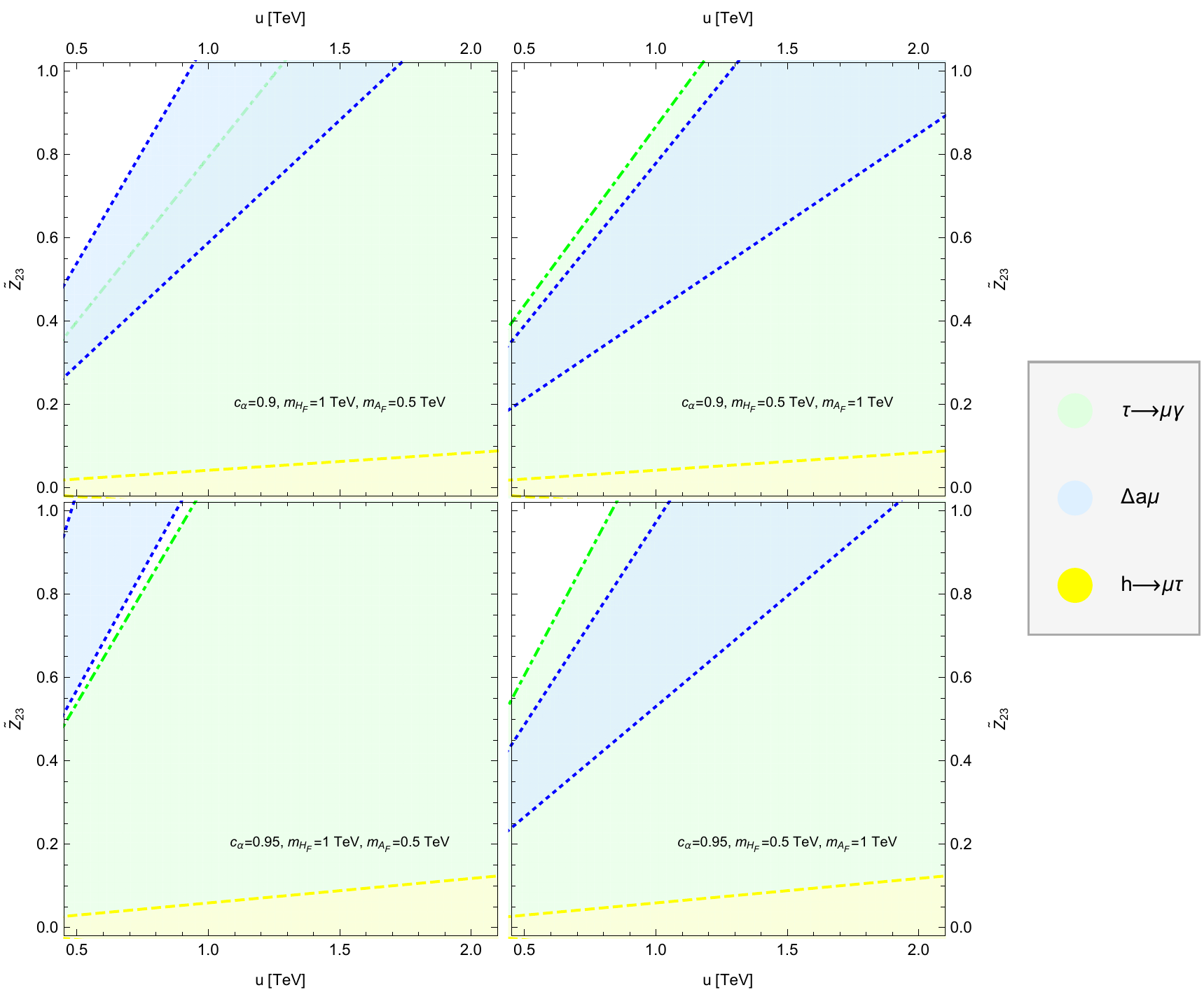}
\caption{Allowed area  on the $u-\tilde Z_{23}^{l}$ plane by the current experimental bounds on the $\tau\to\mu\gamma$ and $h\to\tau\mu$ decays for the indicated values of the model parameters. The blue strip is the region where $u$ and $\tilde Z^l_{23}$ must lie to satisfy  the $a_\mu$ discrepancy.}\label{ParSpace}
	\end{figure}

We will now assess the viability of both scenarios. To avoid large corrections to the diagonal lepton scalar couplings, we   take $\tilde Z_{22}\simeq 10^{-3}$ and $\tilde Z_{33}\simeq 10^{-2}$. We then show in Fig. \ref{ParSpace} the area allowed in the $u-\tilde Z_{23}$ plane by  the experimental constraints on  the $\tau\to\mu\gamma$ and $h\to\tau\mu$ decays for  the indicated values of the  Flavon  masses and  the mixing angle $c_\alpha$. The blue strip is the region where  $u$ and $\tilde Z_{23}$ must lie in order to alleviate the $a_\mu$ discrepancy.    We observe that  the most stringent limits on $\tilde Z_{23}$  are obtained  from the experimental bound on the $h\to \mu \tau$ decay, which  requires   $\tilde Z_{23}$  to be of the order of $10^{-2}$ ($10^{-1}$) for $u=0.5$ ($1$) TeV. However, in order to solve the $a_\mu$ discrepancy,  values of $\tilde Z_{23}$ as large as 1 would be required for $u=1$ TeV, which is due to the fact that we need large (positive) contributions arising from the $CP$-even Flavon and the new corrections to the SM Higgs boson couplings.  We note that the blue strips shifts downwards as the mass of the $CP$-odd Flavon increases as in this case the (negative) contribution is smaller.  Therefore, scenario I discussed above is not favored by the current experimental data and  we assume that the scenario II is fulfilled, with the Flavon contribution to $a_\mu$ being rather small and not responsible for the discrepancy between experimental and theoretical values. We thus take as benchmark $Z_{23}\simeq 0.1$, which is allowed by the  experimental bound on the $\tau\to\mu\gamma$ and $h\to \mu \tau$ decays  for $u\simeq 1.5$ TeV.

\subsection{$\tilde Z^u_{33}$ and  $\tilde Z^u_{23}$ matrix elements}

According to the universal Higgs fit \cite{Giardino:2013bma}, the allowed value for the deviation of the SM $h\bar t t$ coupling is
$\epsilon_t= -0.21\pm 0.22$.   We will use again the two-family
approximation and take  the values
$\tilde{Z}^u_{33}=0.01$ and  $\tilde{Z}^u_{23}=0.1$.

\subsection{Summary of Benchmarks for the model parameters}

In summary, in our study below we will  use the following benchmarks:

\begin{enumerate}

\item {{Mixing angle $\alpha$}}: As discussed above, to satisfy the fit on the 125 GeV Higgs couplings measured at the LHC, the following constraint must be obeyed   $0.86 < c_{\alpha} < 1$. We will thus use the benchmark $c_\alpha=0.95$.

\item {{FN singlet VEV} $u$}: It appears in the  Flavon  couplings and the LFV SM Higgs coupling. We will    consider the value  $1.5$ TeV.

\item {$\tilde{Z}^f_{ij}$ {matrix}}: It determines the strength of the LFV scalar couplings. We will use the 2-family
approximation and take  $\tilde{Z}^l_{22}=10^{-4}$,  $\tilde{Z}^l_{33}=10^{-2}$, and  $\tilde{Z}^l_{23}=10^{-1}$, which are consistent with the constraint on the LFV decay $h\to \mu\tau$.

\item {{$H_Fhh$ interaction:}} This vertex depends on a combination of parameters that appear in the Higgs potential. However, these parameters could be traded by an effective coupling $\lambda_{H_{F}hh}$, which can take values of the  order of $O(1)$. We will thus fix $\lambda_{H_{F}hh}\simeq 0.1 u$.

\end{enumerate}
A summary of the benchmarks we are going to consider in our analysis is presented in Table \ref{Benchmarks}.

\begin{table}
  \centering
  \caption{Benchmarks used for the analysis of the production and detection of the Flavons $H_F$ and $A_F$ at the LHC and a future 100 TeV $pp$ collider in the context of the FNSM.}
  \begin{tabular}{ccc}
    \hline
    \hline
    Parameter
    &Benchmark \\
    \hline
    \hline
    $c_\alpha$  &0.95\\
    $u$&$1.5$ TeV \\
    $\tilde{Z}^l_{22}$ &$10^{-4}$ \\
    $\tilde{Z}^l_{33}$ &$10^{-2}$\\
    $\tilde{Z}^l_{23}$ &$10^{-1}$\\
    $\lambda_{H_F hh}$& $0.1u $\\
    \hline
    \hline
  \end{tabular}
  \label{Benchmarks}
\end{table}

\section{Search for LFV Flavon  decays at hadron colliders}

As stated above, the aim of this work is to analyze the detectability of the LFV signal arising from the Flavon  decays, as predicted
by the FNSM, at the LHC and a future 100 TeV $pp$ collider. Below we will present an analysis concentrating on the
main Flavon production mechanism, i.e. gluon fusion, as well as the branching ratios of its dominant decay modes. Then, we will present the Monte Carlo analysis of the $H_F \to \tau \mu$ and $A_F \to \tau \mu$ decay signatures,
including the study of the potential SM   background. We will present a conservative analysis meant
to find out whether it is possible to have evidence of our signal at the LHC and the future 100 TeV $pp$ collider.

\subsection{Production cross-sections of the $CP$-even and $CP$-odd Flavons}

We now turn to analyze the  main production mode of both  $H_F$ and $A_F$
Flavons at hadronic colliders, namely, by gluon fusion. The High-Luminosity Large Hadron Collider project aims to increase potential discoveries contemplating a luminosity of up to $\mathcal{L}$=3 ab$^{-1}$ about 2025.  As far as a 100 TeV $pp$ collider is concerned the future circular collider (FCC) contemplates an integrated luminosity of until $\mathcal{O}$(10 ab$^{-1}$). We consider the integrated luminosities shown in tha table \ref{tabla lumi}.
\begin{table}

\caption{Integrated luminosities considered in our analysis.\label{tabla lumi}}

\begin{centering}
\begin{tabular}{l r r r}
\hline
Collider & Luminosity\tabularnewline
\hline
\hline
HL-LHC & 0.3-3 ab$^{-1}$\tabularnewline
\hline
FCC & 3-20 ab$^{-1}$\tabularnewline
\hline
\end{tabular}
\par\end{centering}

\end{table}

\begin{figure}[htb!]
\includegraphics[angle=270, width = 7.5cm]{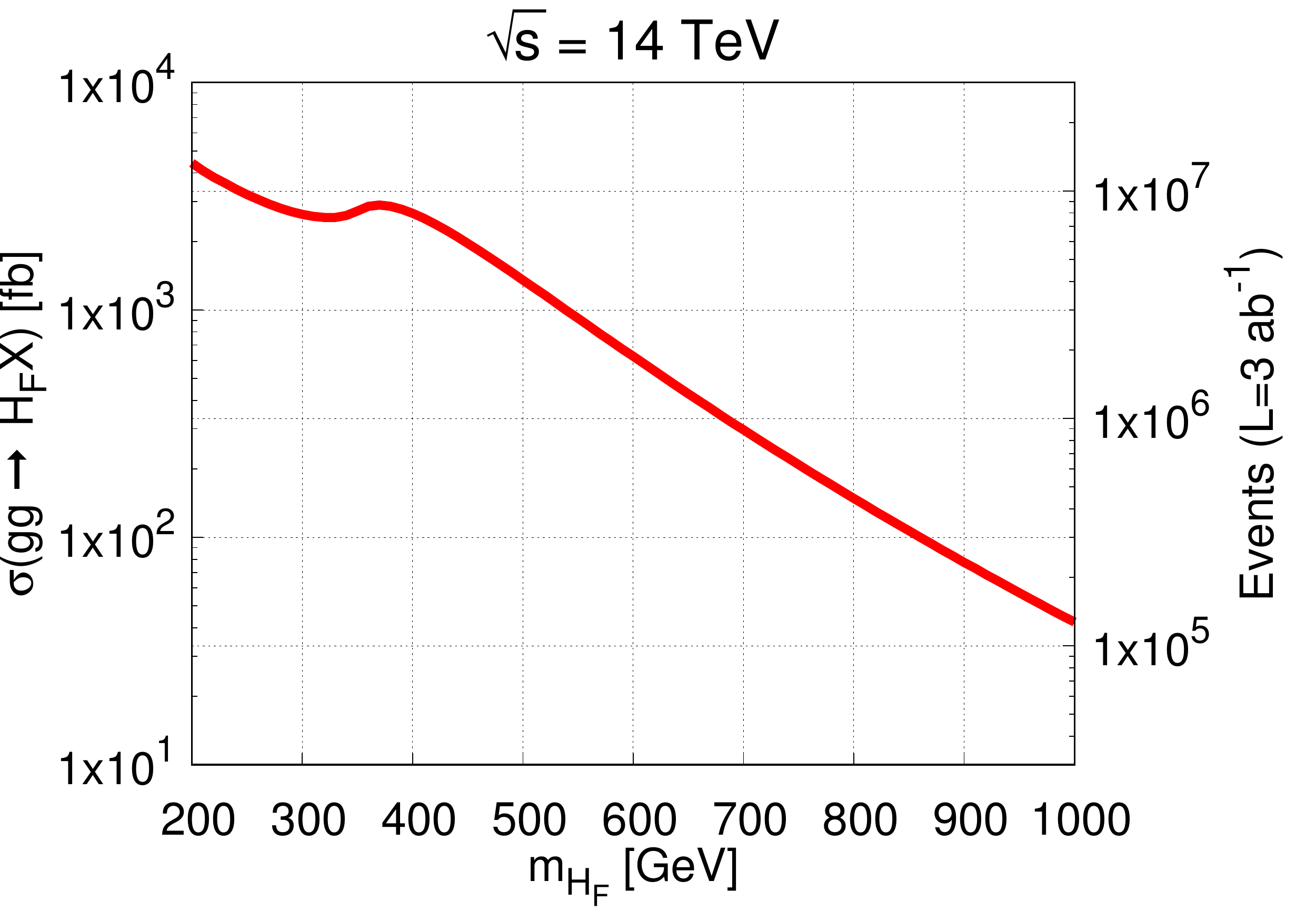}
\includegraphics[angle=270, width = 7.5cm]{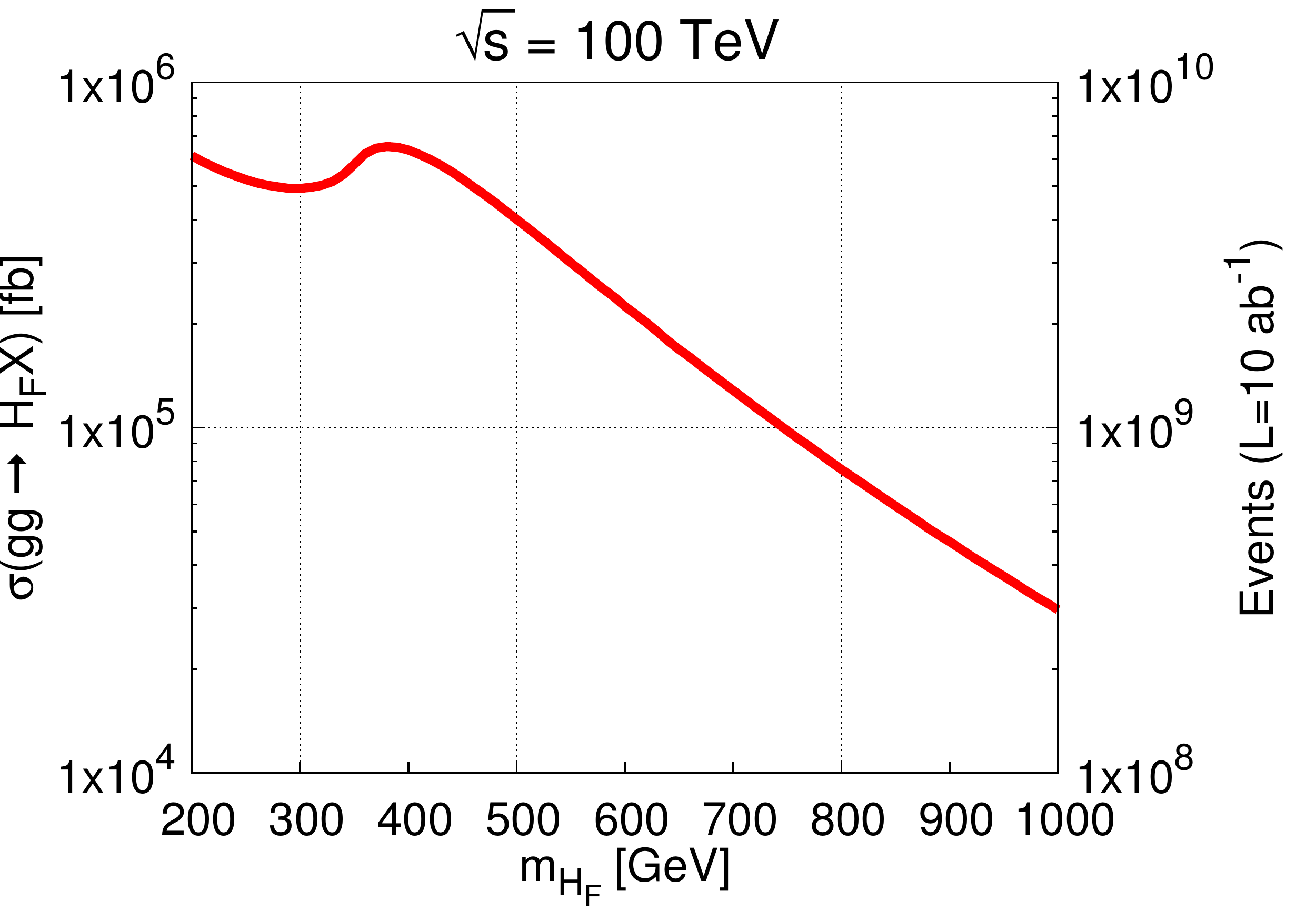}
\caption{Production cross-section $\sigma(pp\to H_F X)$ of  a $CP$-even Flavon  at a hadronic collider as a function of its mass $m_{H_F}$ for $\sqrt{s}=14\;\;(100)$ TeV. The event numbers obtained with an integrated luminosity of  $\mathcal{L}=3\;\;(10)\,\;{\rm ab}^{-1}$ are presented on the right axis.\label{CSProdHF}}
	\end{figure}

\begin{figure}[htb!]
\includegraphics[angle=270, width = 7.5cm]{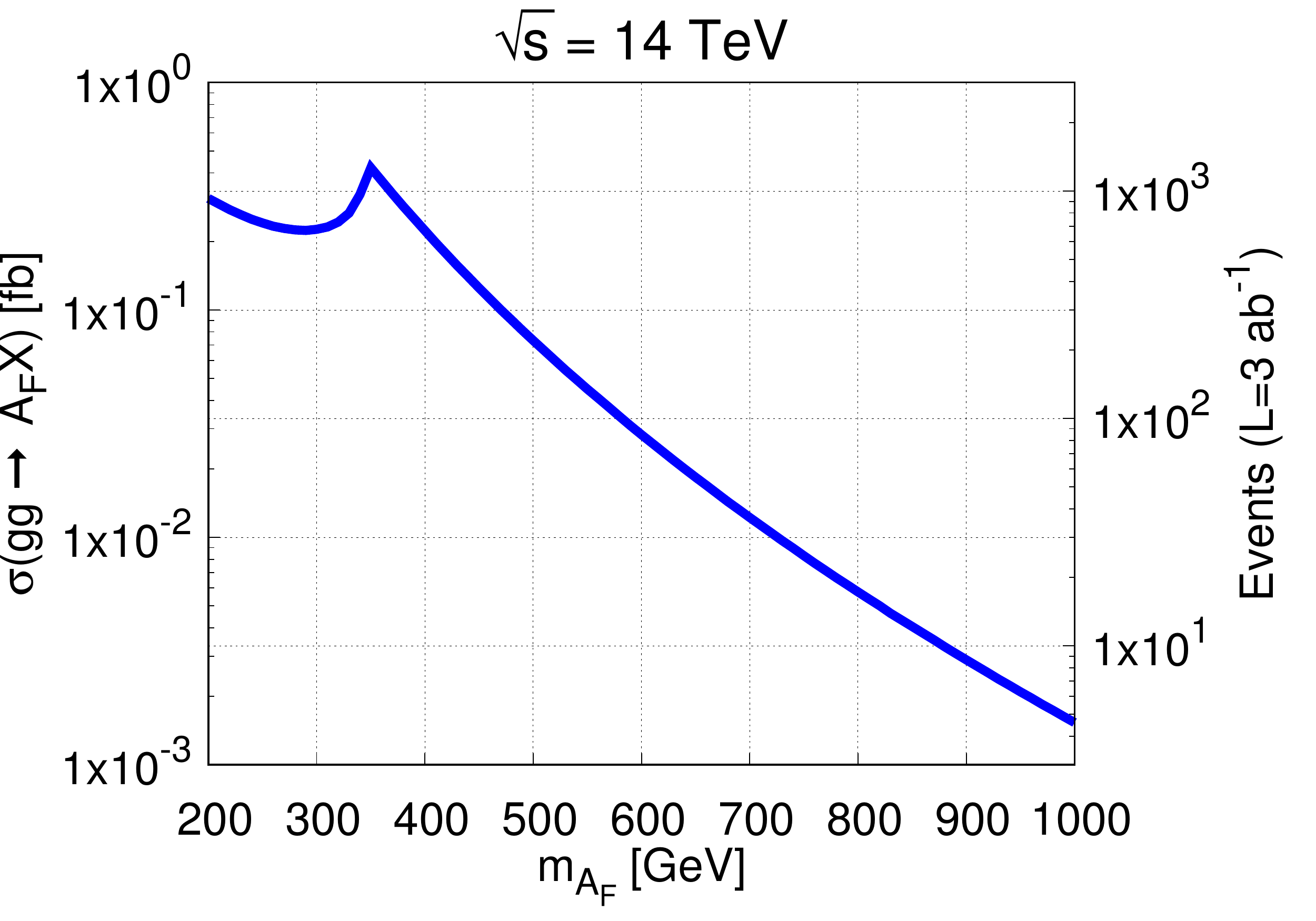}
\includegraphics[angle=270, width = 7.5cm]{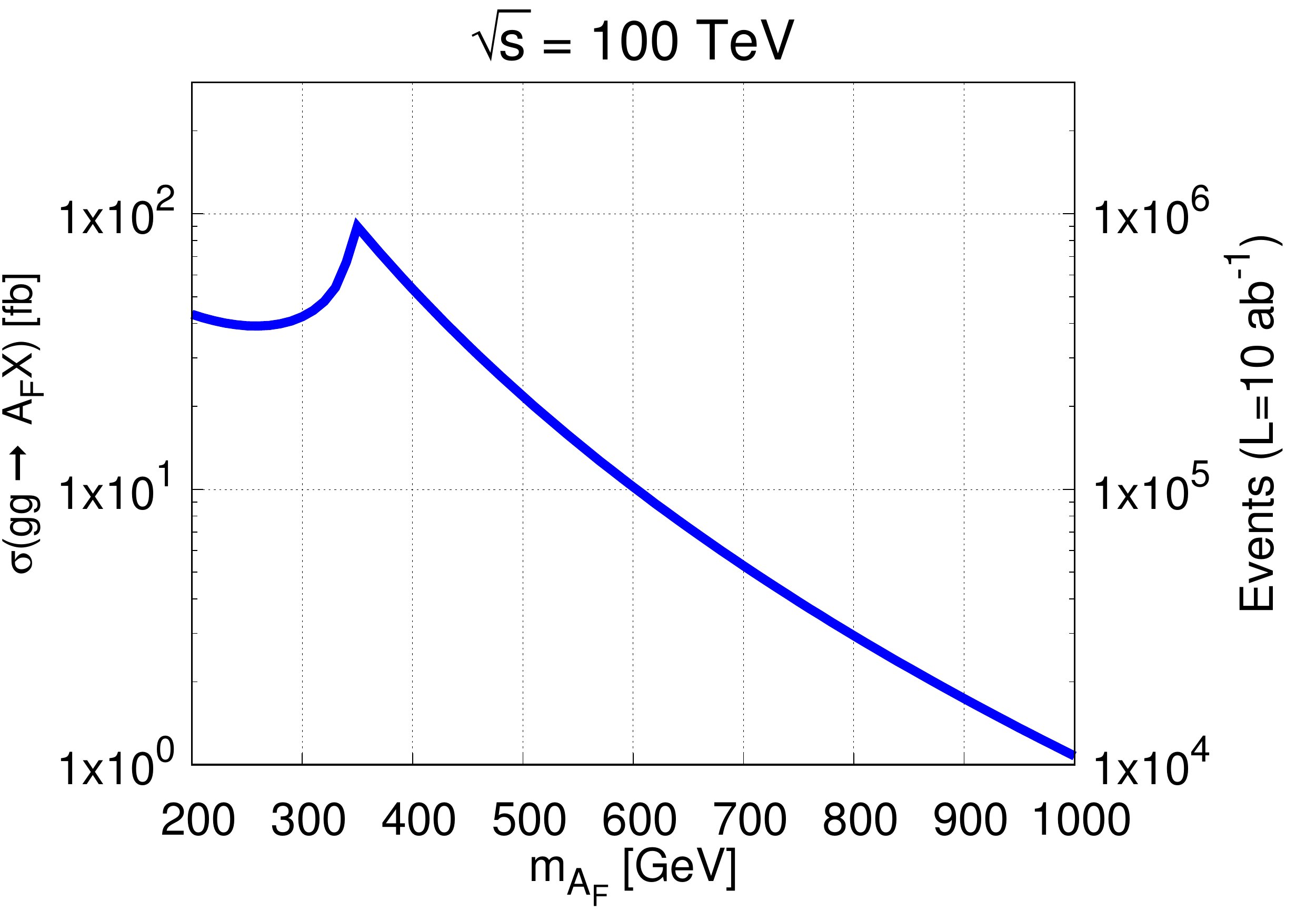}
\caption{The same as in Fig. \ref{CSProdHF} but for a $CP$-odd Flavon . \label{CSProdAF}}
	\end{figure}

In Fig.  \ref{CSProdHF} we show the $pp\to H_FX$ production cross section of a $CP$-even Flavon   as a function of its mass $m_{H_F}$  at the LHC and a future 100 TeV $pp$ collider. We also show the event numbers   on the right axis of each plot.
As far as the   $CP$-odd Flavon  is concerned, the respective production cross section and event numbers  are presented in Fig. \ref{CSProdAF}.
	
\begin{figure}[htb!]
\includegraphics[ width = 6cm]{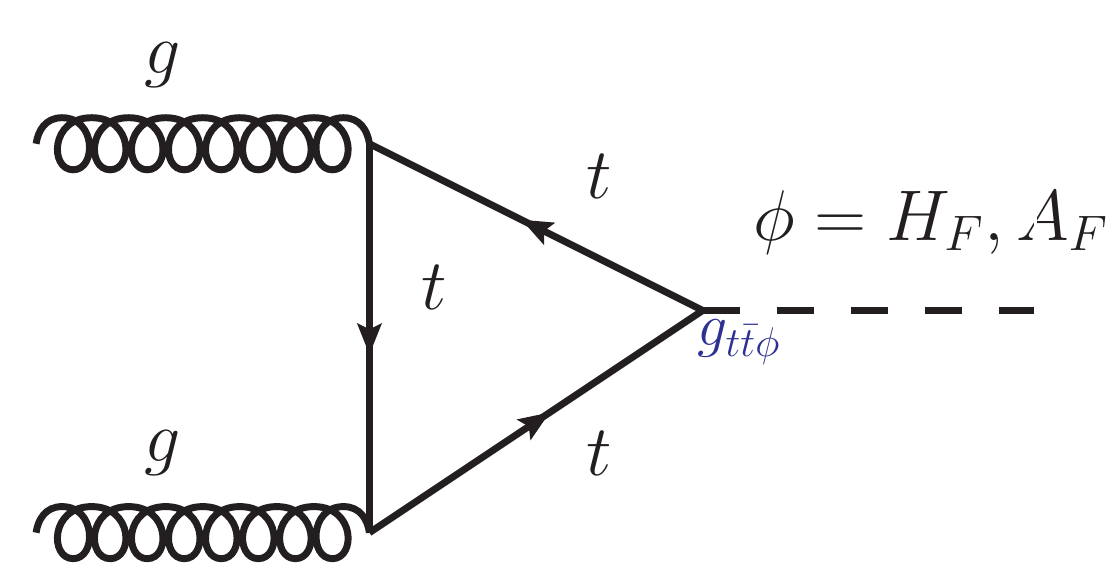}
\caption{Feynman diagram for the dominant contribution to Flavon production via gluon fusion at the leading order. \label{FDgluonfusion}}
\end{figure}

The dominant contribution to gluon fusion arises from loops carrying the top quark as shown in Fig \ref{FDgluonfusion}.  This explains the  suppression of the production of the  $CP$-odd Flavon  as compared to that of the $CP$-even one, as observed in Figs. \ref{CSProdHF}  and \ref{CSProdAF}, which stems from the appearance of  the coupling $g_{A_F\bar{t}t}$ in the corresponding cross section. For instance, taking into account the parameter values of Table \ref{Benchmarks}, we have  $g^2_{A_F\bar{t}t}\sim 10^{-6}$,
whereas for the $CP$-even Flavon  $g^2_{H_F t\bar{t}}\sim10^{-2}$.

\subsection{Flavon  decays}
\subsubsection{Two-body decays}
\begin{figure}[htb!]
\subfigure[ ] {\includegraphics[ width = 5cm,angle=270]{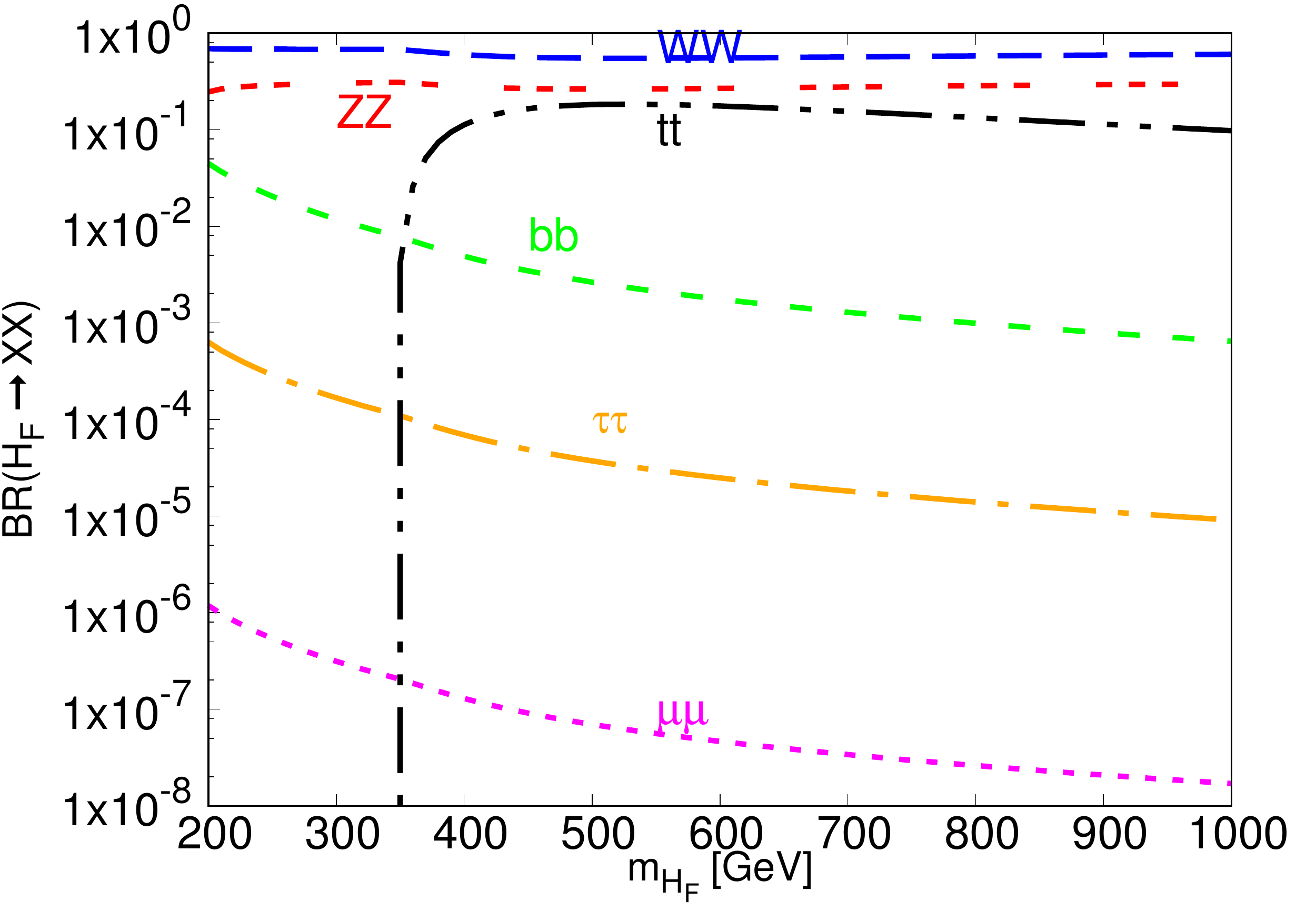} \label{a}}
\subfigure[ ] {\includegraphics[ width = 5cm,angle=270]{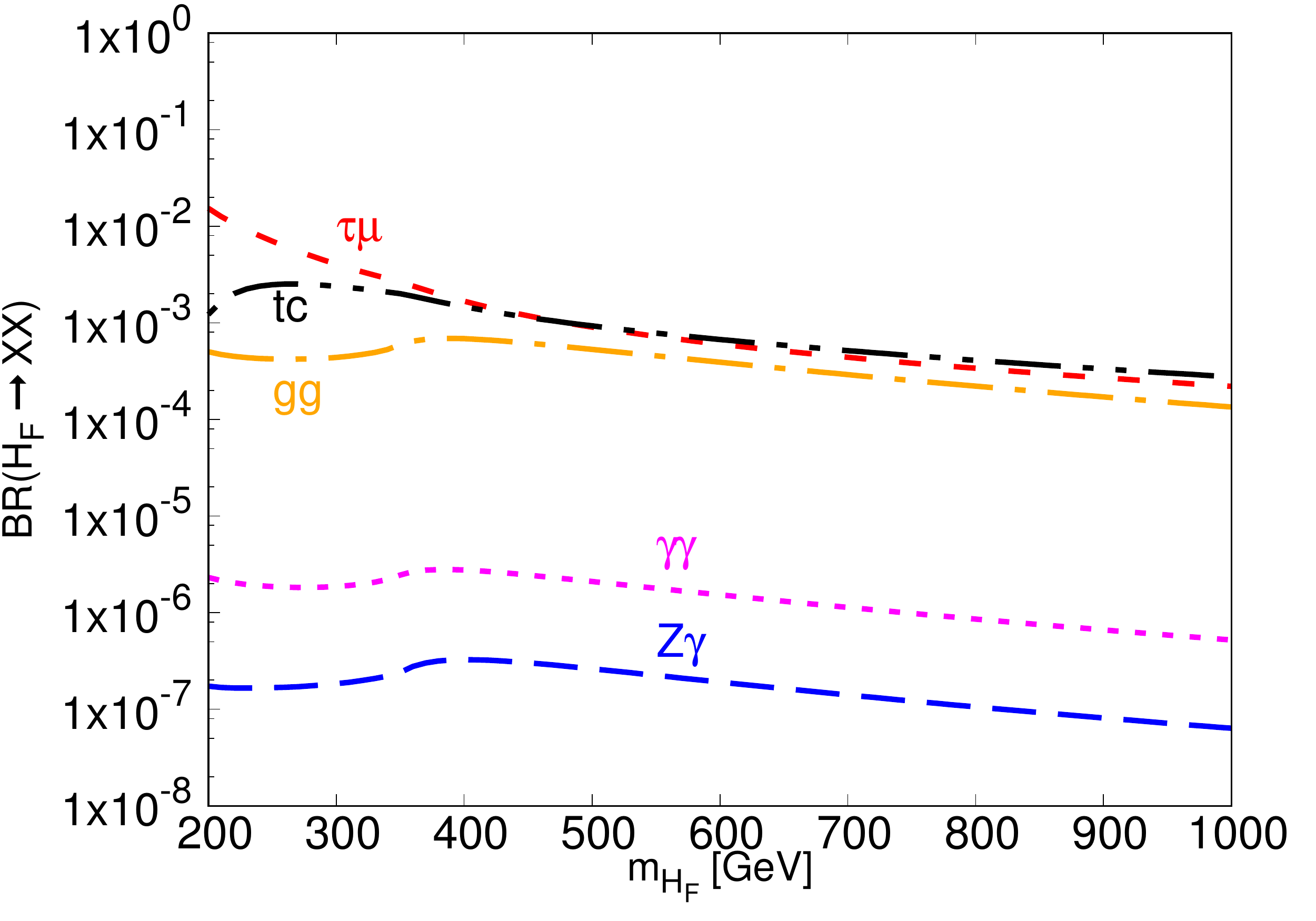} \label{b}}
\caption{Branching ratios of the two-body decay modes of a $CP$-even Flavon    as  a function of its mass for the parameter values of Table \ref{Benchmarks}. \label{BRFlavonHF}}
\end{figure}

We now analyze the behavior of the branching ratios of the dominant  decay modes, including the FV ones, of both the $CP$-even and $CP$-odd Flavons. Analytical expressions for the partial decay widths  are presented in Appendix \ref{DecayWidthFormulas}. It is worth mentioning that a crosscheck was done by comparing  the numerical results obtained via our own C language  code implementing the analytic expressions of Appendix \ref{DecayWidthFormulas},  and  those computed with the aid of  the CalcHEP package \cite{Belyaev:2012qa}, for which we used an implementation of the FNSM Feynman rules obtained with LanHEP \cite{Semenov:2014rea}.  In  Fig. \ref{BRFlavonHF} we present the relevant branching ratios of the decays modes of a $CP$-even  Flavon  as functions of its mass for the benchmarks of  Table \ref{Benchmarks}.

 We observe in Fig. \ref{BRFlavonHF} that in the scenario under study and  for $m_{H_F}$ ranging between $200$ and $1000$ GeV, the dominant $H_F$ decay mode would be  $H_F\to WW$, followed by
 $H_F\to ZZ$. Once the $H_F\to \bar{t}t$ channel became open, its branching ratio would be about the
 same order of magnitude as that of the $H_F\to ZZ$ decay. Other relevant   decay modes would be
 $H_F\to \bar{t}c$, $H_F\to \bar{b}b$, and  $H_F\to \tau\mu$, whereas the one-loop induced decays
 $H_F\to \gamma\gamma$ and $H_F\to \gamma Z$ would have tiny branching ratios.

\begin{figure}[htb!]
\includegraphics[ width = 10cm]{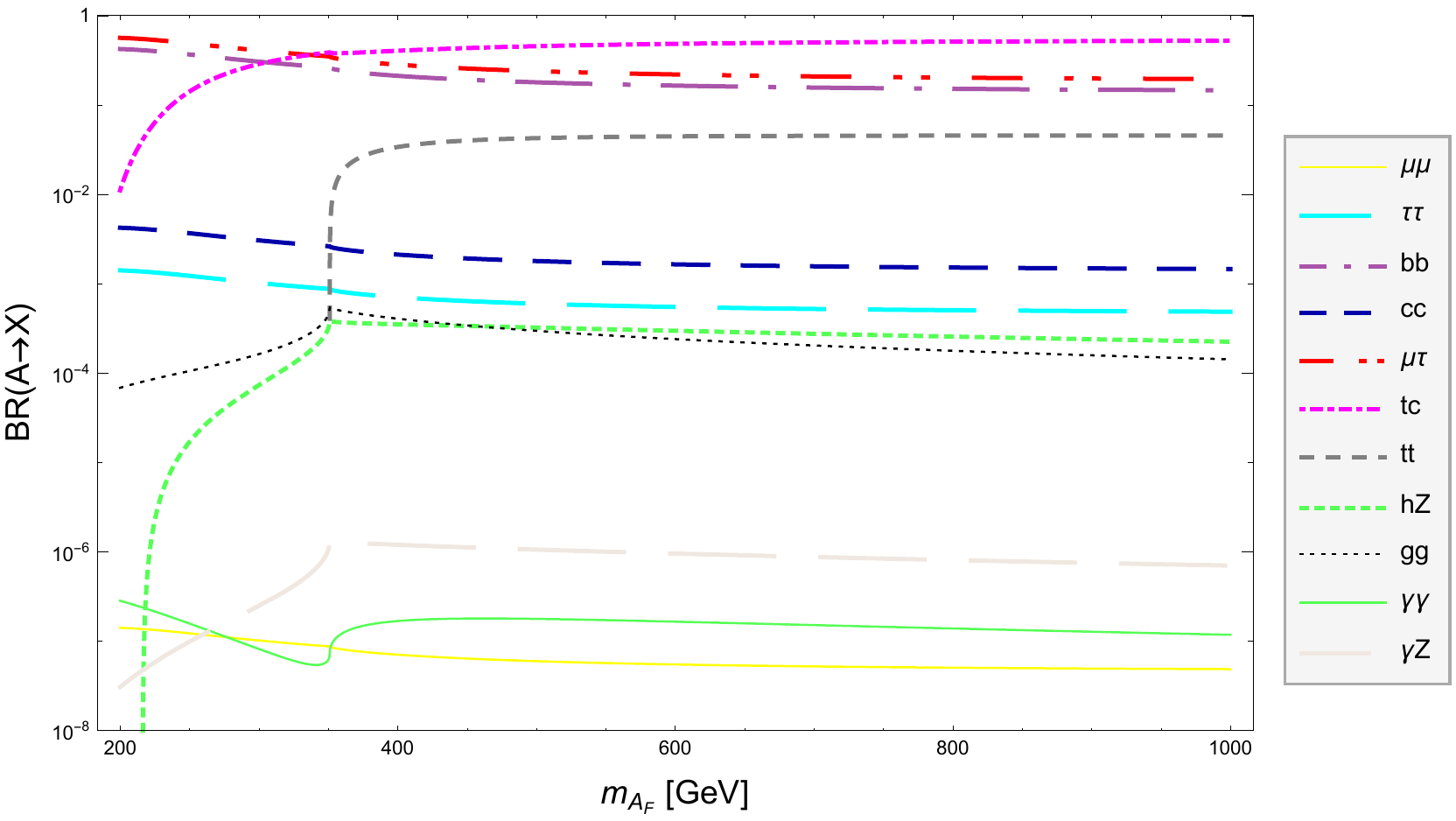}
\caption{The same as in Fig. \ref{BRFlavonHF} but for the $CP$-odd Flavon. \label{BRFlavonAF}}
\end{figure}

As far as the $CP$-odd Flavon  $A_F$ is concerned, since it does not couple to gauge bosons at the tree-level,
its main decay modes are   into fermion pairs. The corresponding branching ratios are shown in
Fig. \ref{BRFlavonAF}.


We can conclude that  the decay modes $A_F\to\bar f f$ could have  branching ratios  up to two
orders of magnitude larger  than the analogue branching ratios of the $CP$-even Flavon .
For the parameter values used here, the main decay channels of the $CP$-odd Flavon  would be
$A_F\to \bar{t}c$, $A_F\to \bar{b}b$, $A_F\to \tau\mu$, whereas the $A_F\to \bar{t}t$ decay would be very suppressed due to the $A_F\bar{t}t$ coupling. As for the one-loop decays $A_F\to \gamma\gamma$ and
$A_F\to \gamma Z$, they  have very small branching ratios,  of the order of  $10^{-8}$.
Along this line,  in Fig. \ref{BRFlavonAF} we also have included the branching ratio for the one-loop induced decay $A_F\to h Z$, which proceeds via the fermion loops shown in Fig. \ref{AtohZ}. We have calculated the corresponding decay width, which is shown in Appendix \ref{DecayWidthFormulas}. It has pointed out recently \cite{Bauer:2016zfj} that this decay could have a relevant branching ratio in models with a $CP$-odd scalar boson arising from a complex singlet and it could be useful to look for evidences of $CP$ violation. We observe in Fig. \ref{BRFlavonAF} that  this decay has a somewhat suppressed branching ratio in the scenario of the FNSM we are interested in: $BR(A_F\to h Z)\simeq 10^{-4}$ for $m_{A_F}$ in the interval between 200 GeV  and 1000 GeV. This value is much larger than the branching ratios of the loop induced decays $A_F\to \gamma Z$ and $A_F\to \gamma\gamma$  but it is about the same size than $BR(A_F\to gg)$. A rough estimate shows that other loop induced decays such as $A_F\to ZZ$ and $A_F\to WW$ have also  suppressed branching ratios. This stems from the fact that  not only they are of higher order in the coupling constants but are  suppressed by the loop factor   $1/(16 \pi^2)$, which enters squared into the decay width. In general, the one-loop induced branching ratios $BR(A_F\to VV)$ ($V=W,Z$) and $BR(A_F\to hZ)$  are suppressed with respect to those of the tree-level induced decays $A_F\to \bar{f}f$ by a factor of $(3 g^2 m_t/(16\pi^2m_W))^2\simeq 3.3 \times10^{-4}$, assuming that the main contribution arises from the top quark. Although these one-loop induced decays may be interesting by themselves, they have no impact on our study of the lepton flavor decays of the $CP$-odd Flavon.

\begin{figure}[!htb]
  \centering
  \includegraphics[width=10cm]{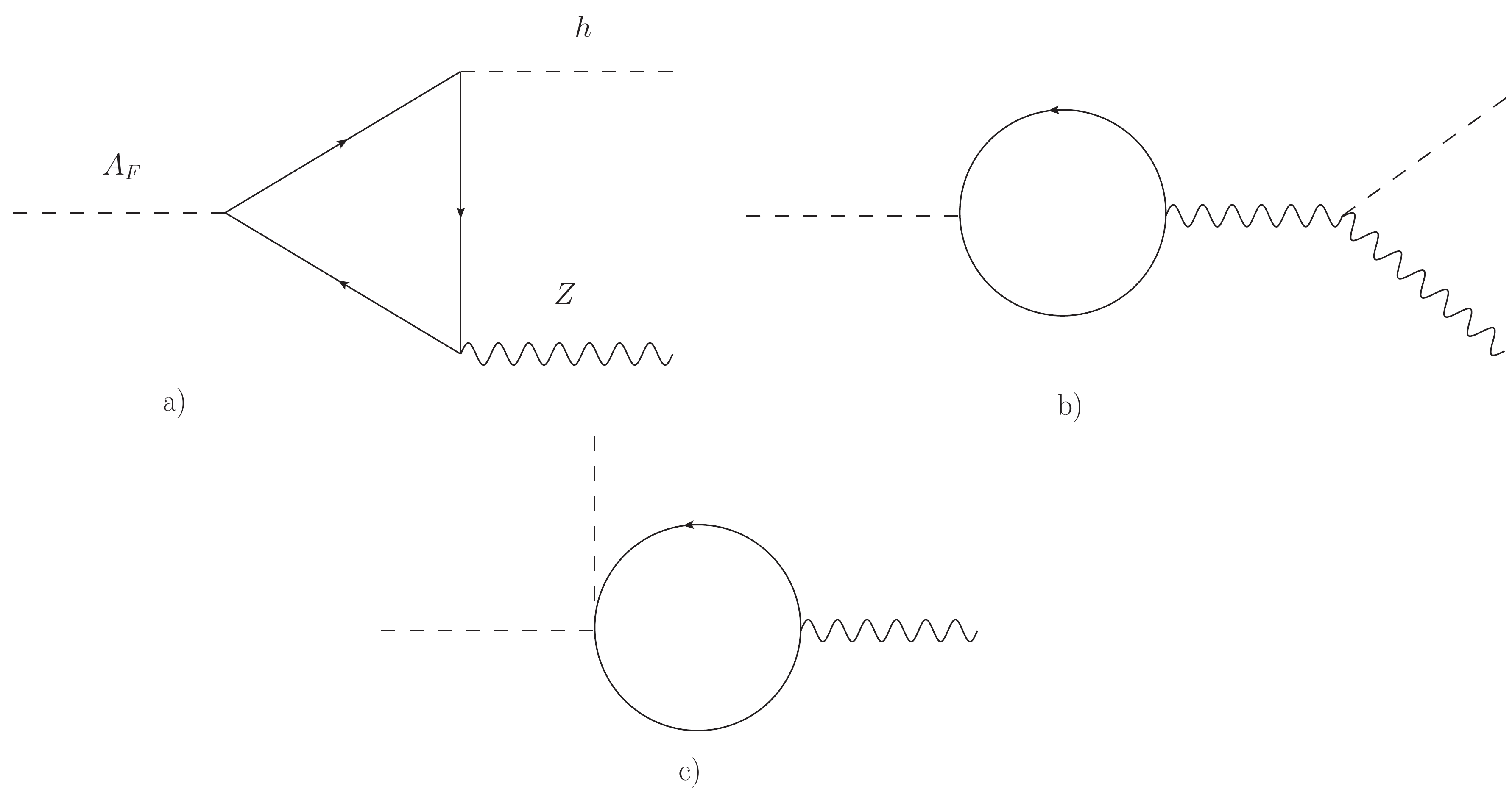}
  \caption{Fermion loop contribution to the $A_F\to hZ$ decay in the FNSM, where it is absent at the tree-level. There is one additional triangle diagram where $h$ and $Z$ are exchanged. We denote the four-momenta as follows $A_F(p)\to h(p_1) Z(p_2)$ ($p$ is incoming whereas $p_1$ and $p_2$ are outgoing). }\label{AtohZ}
\end{figure}

 \subsubsection{Three-body decays $H_F \to \bar{f}_i f_j h$}
\begin{figure}[htb!]
\centering
\addtocounter{subfigure}{0}
\subfigure[ ]{\includegraphics[angle=270, width = 7cm]{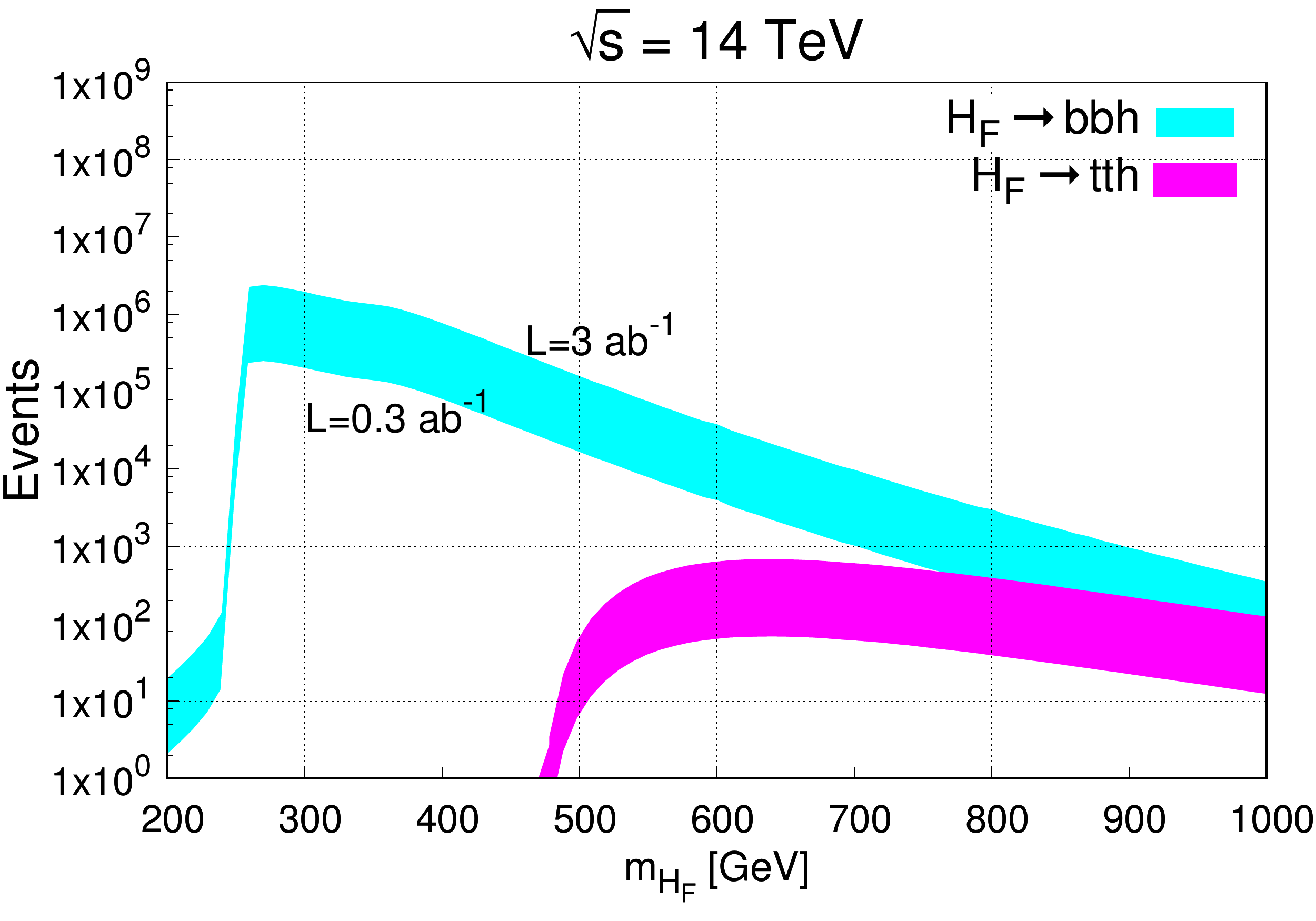}}
\subfigure[ ]{\includegraphics[angle=270, width = 7cm]{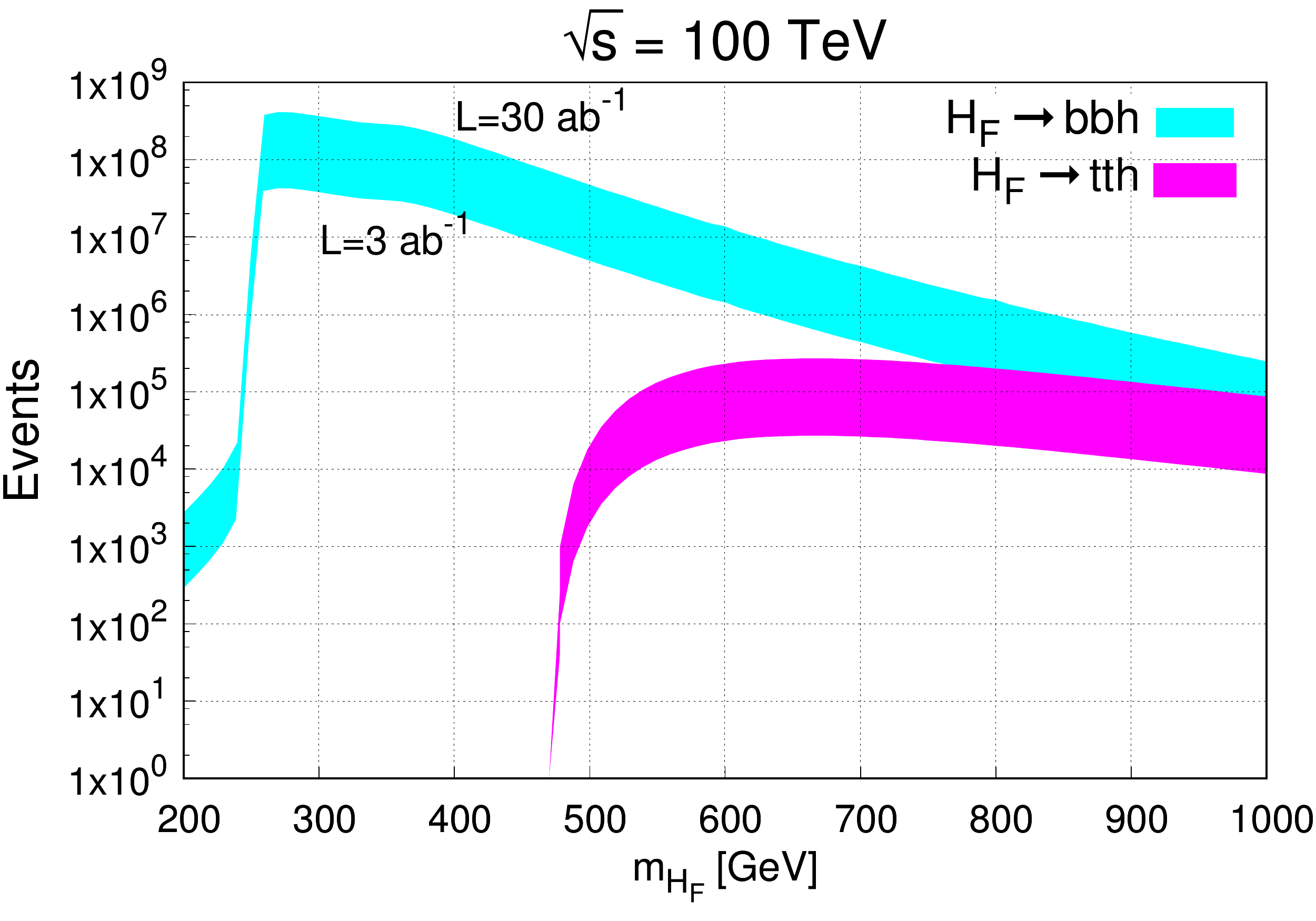}}
\caption{Event number for the  process $pp\to H_F\to q\bar{q}h$ ($q=b,\,t$)  as a function of $m_{H_F}$: (a) $\sqrt{s}=14\;\;$TeV for an integrated luminosity from $\mathcal{L}$=0.3 ab$^{-1}$ (lower limit) to 3 ab$^{-1}$ (upper limit) and (b) $\sqrt{s}=100\;\;$TeV for an integrated luminosity from $\mathcal{L}$=3 ab$^{-1}$ (lower limit) to 30 ab$^{-1}$  (upper limit). The parameters of Table \ref{Benchmarks} were used.}\label{EventsHffh}
\end{figure}

At the tree-level the $CP$-even Flavon  also couples with a fermion pair and a SM Higgs boson, thus
it is worth analyzing the behavior of the three-body decay modes $H_F \to \bar{f}_i f_j h$.
The corresponding decay width is presented in Appendix \ref{DecayWidthFormulas}.
In Fig. \ref{BRHffh} we show the branching ratios for these three-body decays. We observe that, for  a  relatively
light Flavon with  mass around 300 GeV, the decay channel  $H_F\to \bar{b}bh$ would have a branching ratio as large as
$10^{-1}$. On the other hand, other kinematically allowed $H_F \to \bar{f}_i f_j h$  decays  would reach branching ratios as high as $10^{-2}$. For a heavier Flavon with $m_{H_F}>600$ GeV, the decay $H_F\to \bar{t}th$ would become open and   could  be the dominant three-body decay mode  for $m_{H_F}\simeq 900$ GeV. Although these  decay channels seem worth a more detailed study,  we will content ourselves with obtaining the event numbers that could be achieved at the LHC and the FCC. We consider values for the luminosities of the table \ref{tabla lumi}.
In the Fig. \ref{EventsHffh} we shown  the number of events for the processes
$pp\to H_F\to q\bar{q}h$ with $q=b,\,t$.

\begin{figure}[htb!]
\includegraphics[angle=270, width = 8.5cm]{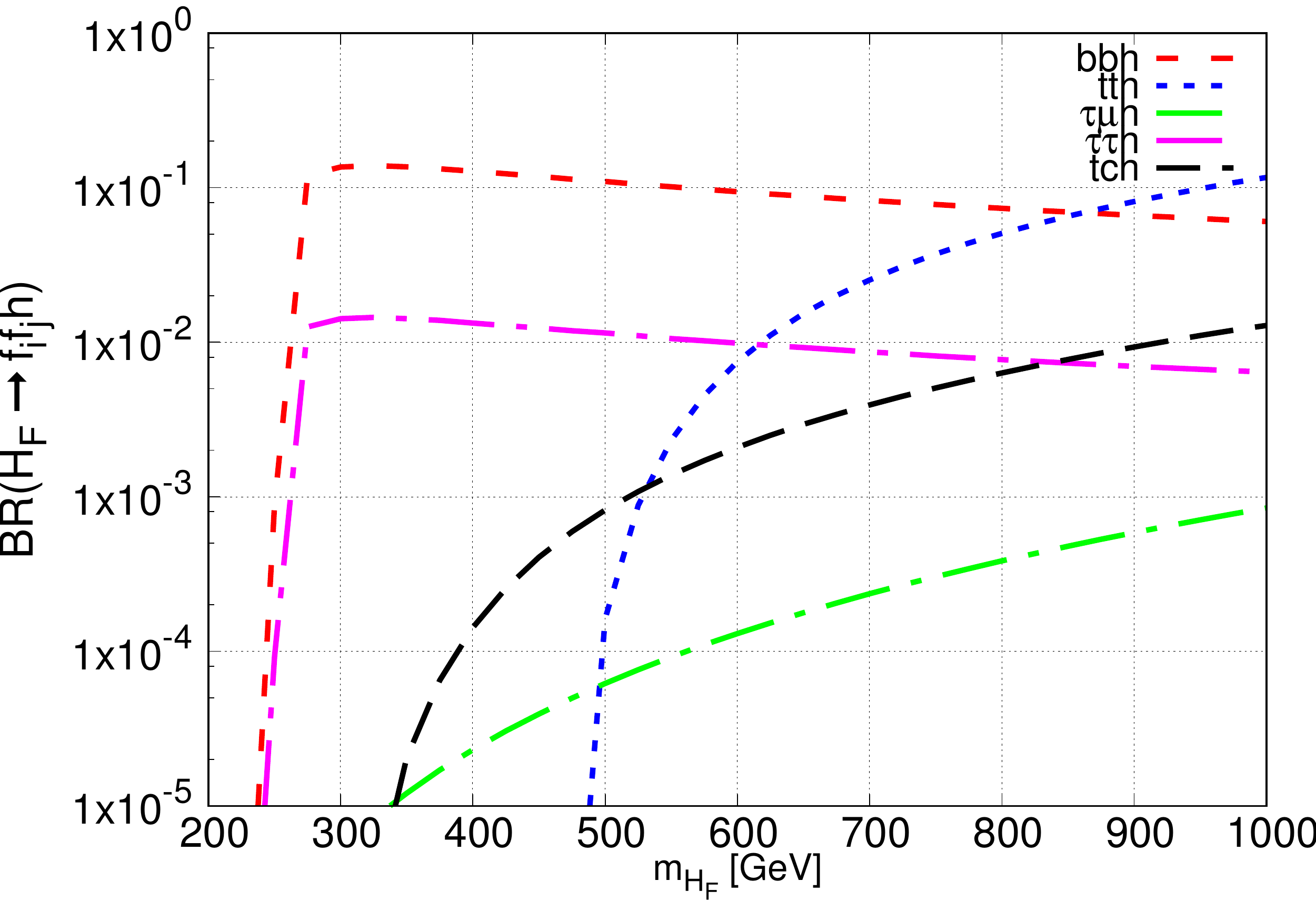}
\caption{Branching ratios for the three-body decay modes $H_F\to \bar{f_i}f_j h$ as  functions of $m_{H_F}$ for the parameters of Table \ref{Benchmarks}.}\label{BRHffh}
\end{figure}

\subsection{Search for LFV Flavon  decays at the LHC and a future 100 TeV $pp$ collider}

\begin{figure}[htb!]
\includegraphics[angle=270, width = 7.5cm]{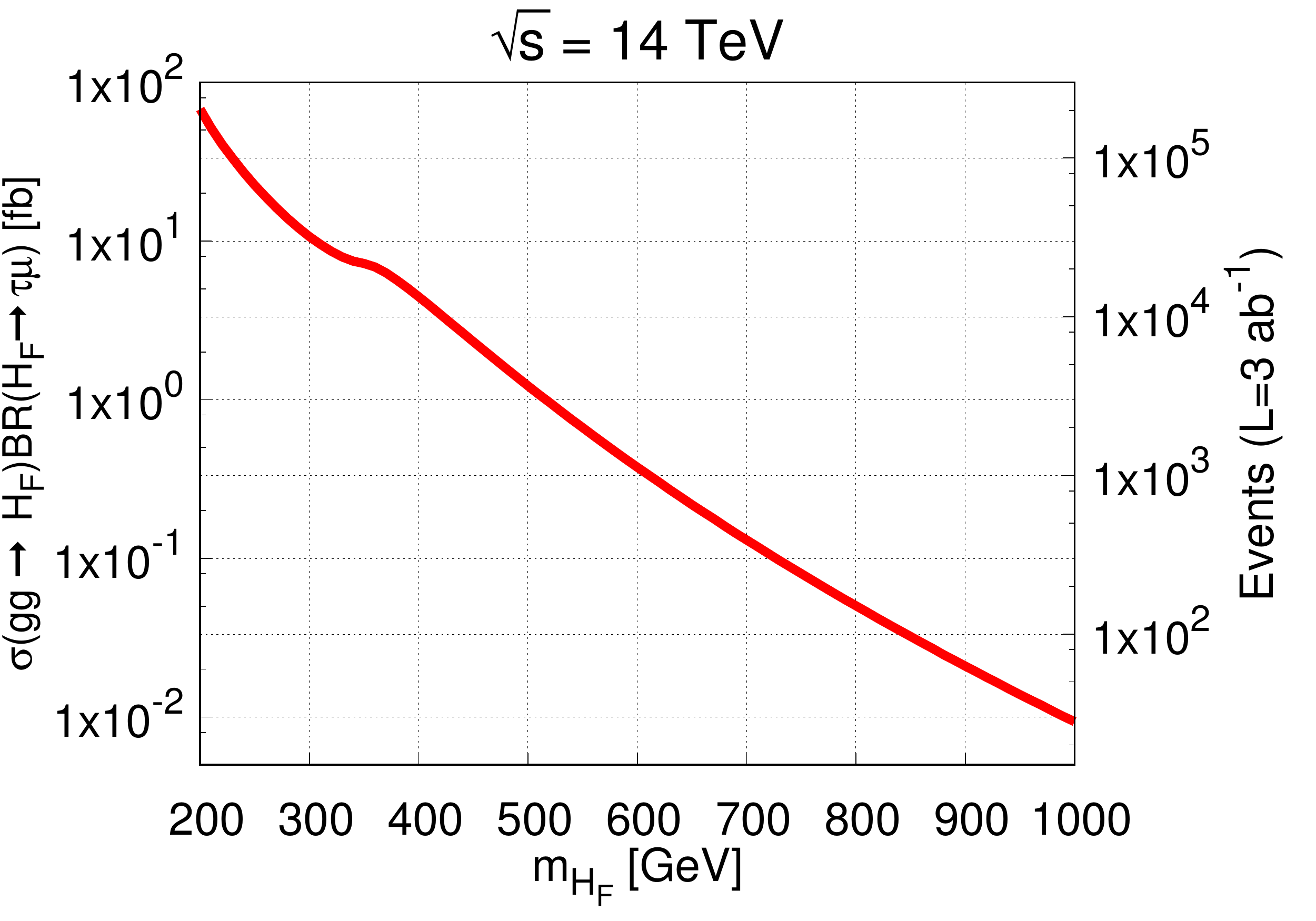}
\includegraphics[angle=270, width = 7.5cm]{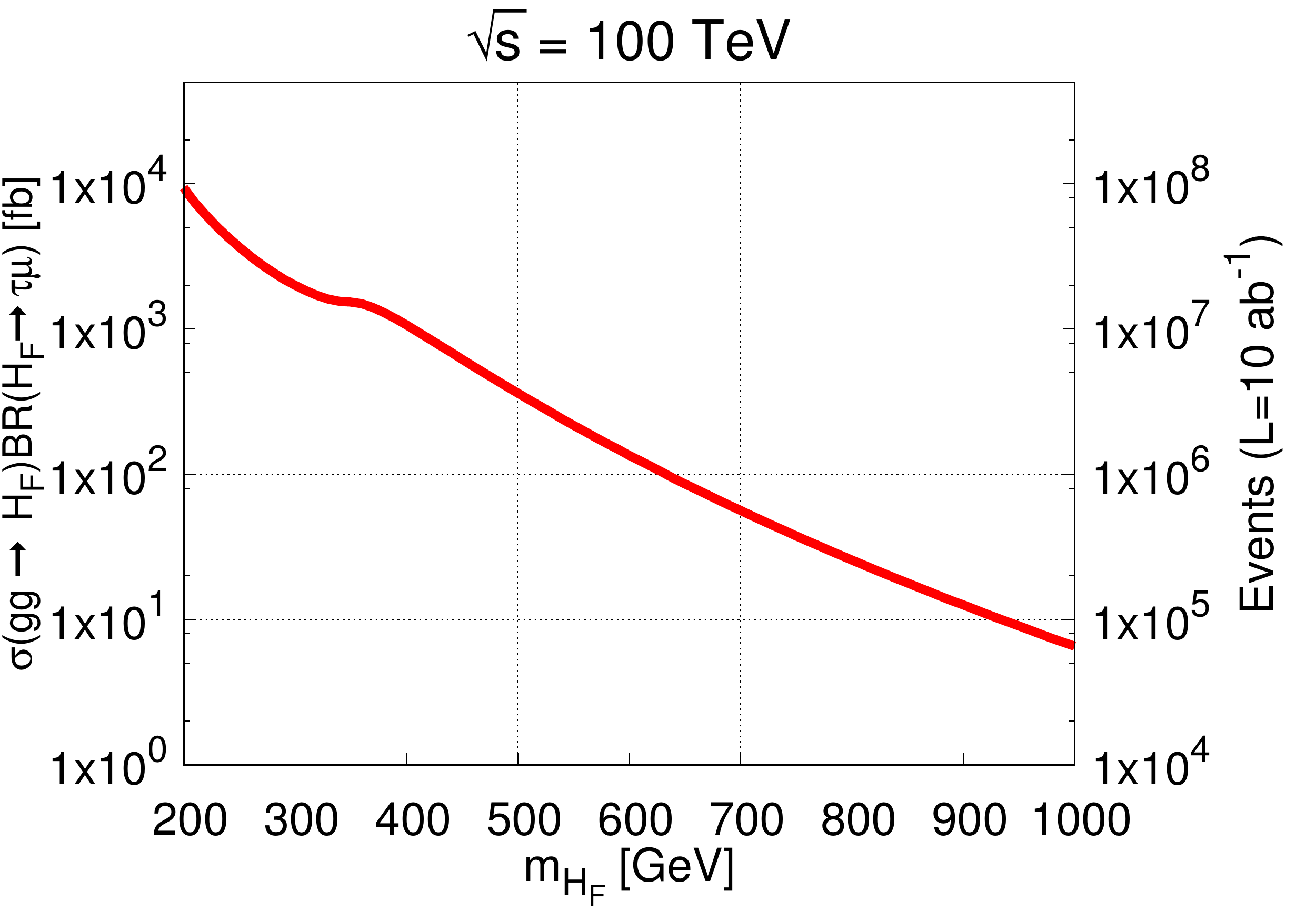}
\includegraphics[angle=270, width = 7.5cm]{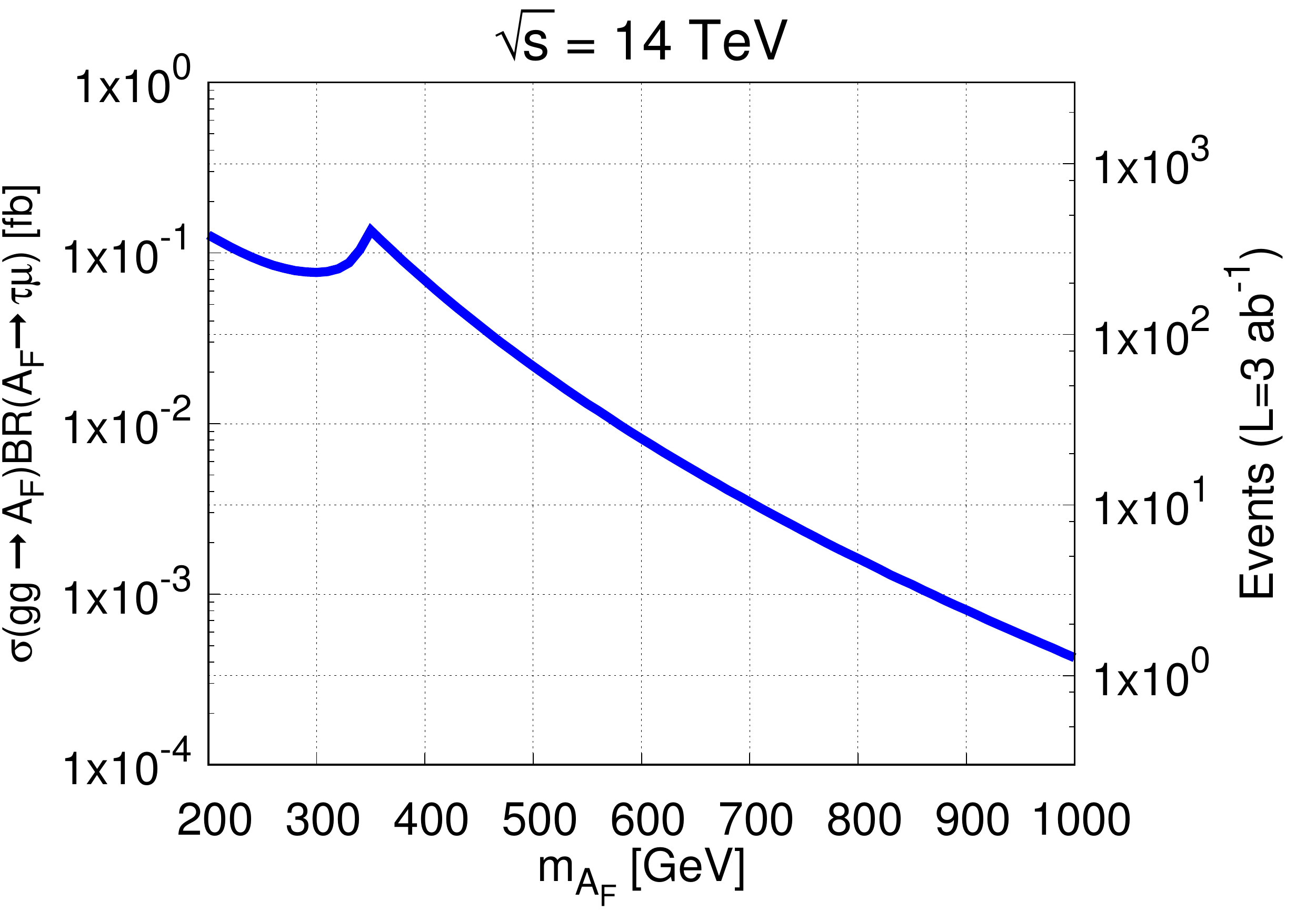}
\includegraphics[angle=270, width = 7.5cm]{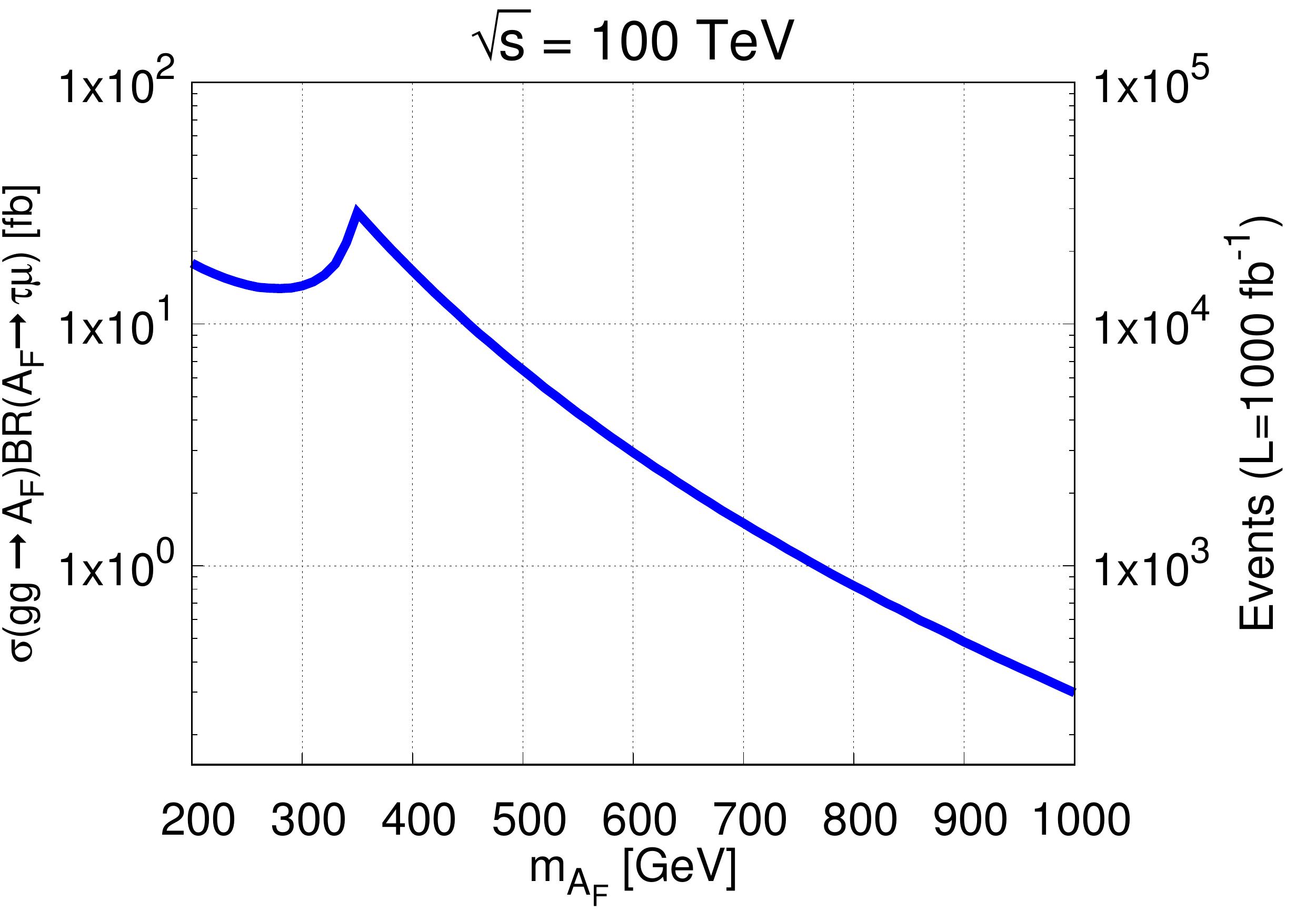}
\caption{$\sigma(pp\to H_F\to \tau\mu)$ (top plots) and $\sigma(pp\to A_F\to \tau\mu)$ (bottom plots) cross sections as functions of the Flavon  mass for $\sqrt s=14$ TeV (left plot) and  $100$ TeV (right plots). On the right axes we show the event numbers considering an integrated luminosity $\mathcal{L}$ of $3$ ab$^{-1}$ and  $10$ ab$^{-1}$. \label{NumEventsSGN}}
	\end{figure}

We are interested in the possible detection of the $CP$-even and $CP$-odd Flavons via their LFV decay into  a $\tau \mu$ pair at the LHC and the FCC. We thus show in Fig. \ref{NumEventsSGN}  the event numbers
for the processes $pp\to H_F\to\tau\mu$ and $pp\to A_F\to\tau\mu$, for  $\sqrt{s}=14\;\;(100)$ TeV and an integrated luminosity for the values displayed in the Table \ref{tabla lumi}. We also use the same set of parameter values of Table \ref{Benchmarks}. We note that for a $CP$-even
Flavon  with a mass  about $1$ TeV, there would be about $\mathcal{O}$(10) $\mu \tau$  signal events at the LHC and $\mathcal{O}(10^4)$
events at the FCC. These event numbers would decrease by about two orders of magnitude
for a $CP$-odd Flavon.

We will now analyze the signature of the LFV Flavon decays  $H_F\to \tau\mu$ and $A_F \to  \tau \mu$,  with $\tau\mu=\tau^- \mu^+ +\tau^+\mu^-$, and their potential SM background. We are inspired in the analysis carried out by the CMS
collaboration in Refs. \cite{Khachatryan:2015kon}, ATLAS collaboration \cite{Aad:2016blu} and the work of the authors of \cite{Primulando:2016eod}. The ATLAS and CMS collaborations considered the two following tau decay channels:
electron decay $\tau_e \mu$ and the hadron decay $\tau_h \mu$. For our analysis we will concentrate instead  on the electron decay. As far as our computation scheme is concerned, we first use the  LanHEP routines to obtain the FNSM  Feynman rules for Madgraph \cite{Alwall:2011uj}. In this way, the signal and background events can be generated
by  MadGraph5 and MadGraph5$\_$aMC$@$NLO, respectively, interfaced with Pythia 6 \cite{Sjostrand:2006za} and Delphes 3 \cite{deFavereau:2013fsa} for detector simulations. The background events were generated at NLO in QCD and the signal at LO, using the  CT10 parton distribution functions \cite{Gao:2013xoa}. The signal and main background events are as follows:

\begin{itemize}
\item \textbf{Signal}: the signal is $gg\to\phi\to\tau\mu\to e \nu_e \nu_{\tau} \mu$ with $\phi=H_F,\,A_F$.  The electron channel must contain exactly two opposite-charge leptons, one an electron and the other a muon. Then, we search for the final state $e\mu$+miss energy. We consider the specific case for which an integrated luminosity $\mathcal{L}$ of $1$ ab$^{-1}$ for the LHC is considered and in the interval $1-20$ ab$^{-1}$ for the FCC.
\item \textbf{Background}: the main SM background arises from $Z$ production via the Drell-Yan process,
followed by the decay $Z \to\tau\tau $ as well as  $W^+W^-$ and $ZZ$ pair production and jets.
 In this work we will only consider the main
background to assess how our signal could be searched for.
\end{itemize}

\subsubsection{Analysis at the LHC}

\begin{figure}[htb!]
\includegraphics[width = 10cm]{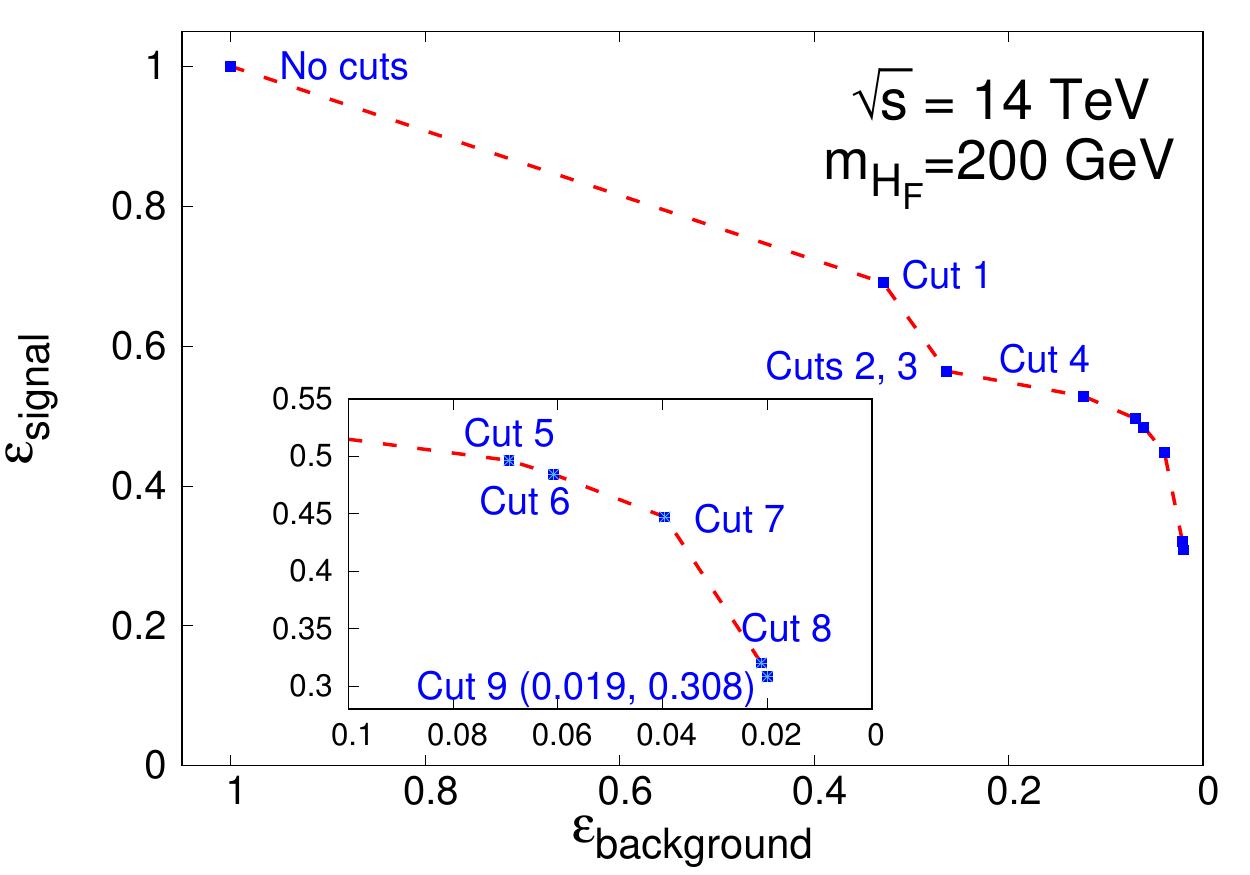}
\caption{Evolution of the efficiencies of the $pp\to H_F\to \tau\mu$ signal and the SM background after the kinematic cuts of Table \ref{CutsLHCHF}  are successively applied. We use  $m_{H_F}=200$ GeV and the parameters of Table \ref{Benchmarks}.\label{Eficiencia}}
\end{figure}

\begin{figure}[htb!]
\includegraphics[angle=270, width = 10cm]{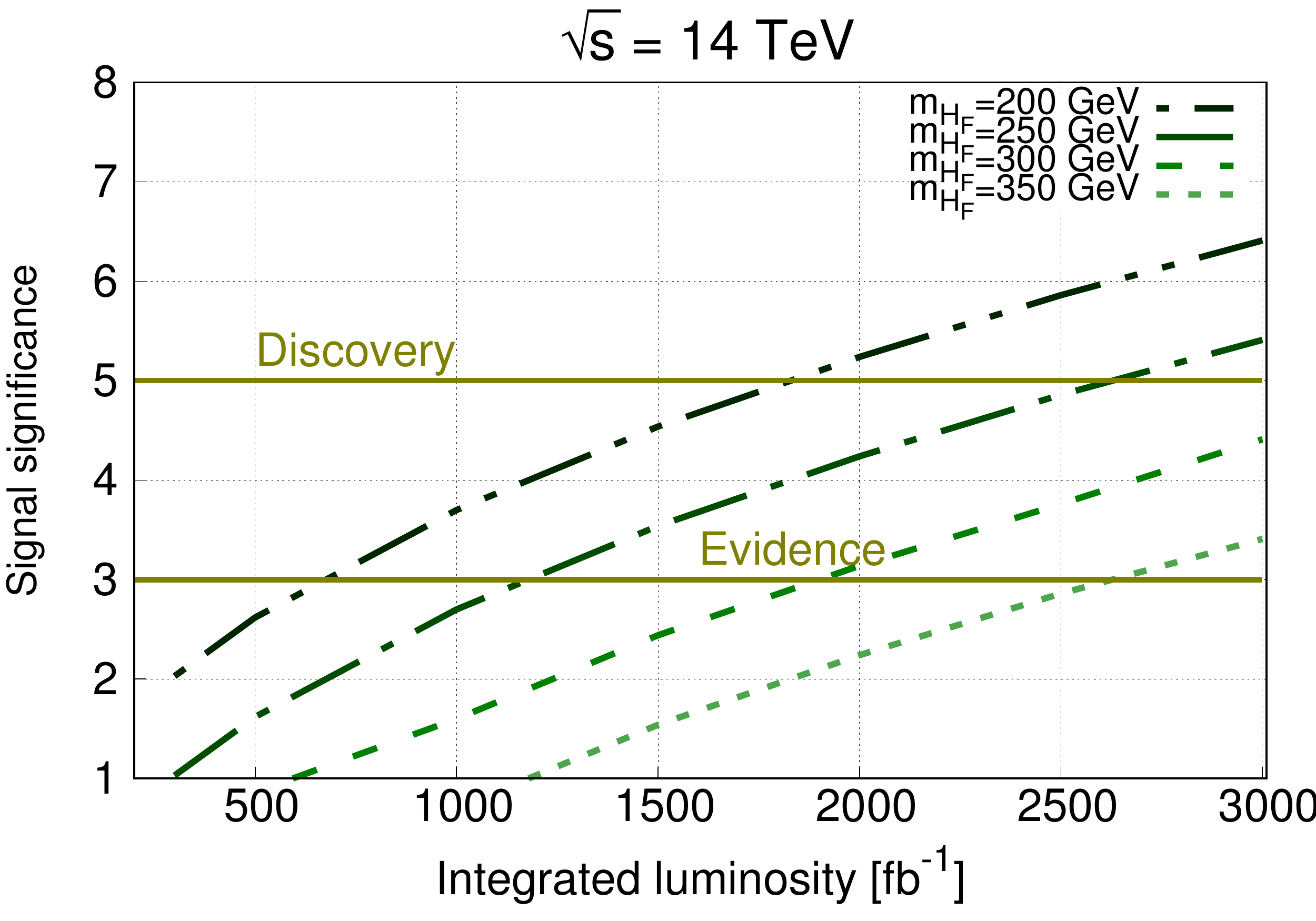}
\caption{Signal significance $S/\sqrt{S+B}$ as a function of the integrated luminosity for the LFV Flavon $\tau\mu$ decay at the LHC. The horizontal lines show the discovery and evidence thresholds. }\label{sigmaVSLumi}
\end{figure}

We start  by analyzing the  possible detection of the $CP$-even Flavon at the LHC  with  $\sqrt{s}$=14 TeV. For illustrative purpose, we use the following set of values  for  $m_{H_F}$: 200, 250, 300, and 350 GeV.
We generated $10^5$ events for  the signal  and the SM main background.
Afterwards, the kinematic analysis was done via MadAnalysis-5 \cite{Conte:2012fm}.
The  cuts applied to both the signal and   background are shown in Table \ref{CutsLHCHF}, where we also show  the  event numbers of the signal (S) and background (B) after the kinematic cuts are applied, along with the signal significance $S/\sqrt{S+B}$ for $m_{H_F}=200$ GeV. The effect of the cuts on the   signal and background event numbers is best illustrated  in Fig. \ref{Eficiencia}, where we show how the  efficiencies $\epsilon_{\text{signal}}$  and $\epsilon_{\text{background}}$ evolve after each cut is successively applied. One can observe that once the kinematic cuts are applied, the resulting signal efficiency is about $0.31$,
whereas that for the background is around  $0.02$. With a luminosity of $0.3$ ab$^{-1}$, the signal
significance is about $2\sigma$. However, if we take into account an integrated luminosity of $3$ ab$^{-1}$ it increases up to $\sim$6.5$\sigma$. The net effect is shown
in Fig. \ref{sigmaVSLumi}, where the signal significance is plotted as a function of the luminosity for the chosen values of $m_{H_F}$.

\begin{table}
\caption{Kinematic cuts applied to the $pp\to H_F\to \tau\mu\to e\mu+\text{miss energy}$ signal and the SM background at the LHC with a center-of-mass energy $\sqrt{s}=14$
TeV and an integrated luminosity of $\mathcal{L}=1$ ab$^{-1}$. We also show the corresponding event numbers obtained after the kinematic cuts and the signal significance $S/\sqrt{S+B}$ for $m_{H_F}=200$ GeV. The tranverse mass is defined as $M_T^{\ell}=\sqrt{2 P_T^{\ell}E_T^{\text{miss}}(1-\text{cos}\Delta\phi_{P_T^{\ell}-E_T^{\text{miss}}})}$ . \label{CutsLHCHF}}
\centering{}%
\begin{tabular}{|c|c|c|c|c|}
\hline
 {Cut number}&\bf{Cuts} & \bf{Signal (S)} & \bf{Background (B)} & $\frac{S}{\sqrt{S+B}}$\tabularnewline
\hline
\hline
&Initial (no cuts) & 9190 &  29320240 & 1.7 \tabularnewline
\hline
1& $|\eta^e|<$2.3&          6348 &          9644078 &          2.04  \tabularnewline
\hline
 2&$|\eta^{\mu}|<$2.1&          5185 &          7736476 &          1.86 \tabularnewline
\hline
 3& 0.1$<\Delta$R($e,\,\mu$)&          5185 &          7727929 &          1.87 \tabularnewline
\hline
 4& 60$<P_T^{\mu}$&          4856 &          3602928 &          2.56 \tabularnewline
\hline
 5& 20$<P_T^e$&          4562 &          2031748 &          3.20\tabularnewline
\hline
 6& 20$<M_{\text{inv}}(e,\,\mu)<$170 &          4450&          1781998 &          3.20  \tabularnewline
\hline
 7& 10$<MET<$100&          3504&          1158653 &          3.33 \tabularnewline
\hline
8& 75$<M_T^{e}$ & 2942 & 973054 & 3.55\tabularnewline
\hline
9& 60$<M_T^{\mu}$ & 2833 & 585342	&3.7\tabularnewline
\hline

\end{tabular}
\end{table}

\begin{figure}[htb!]
\includegraphics[angle=270, width = 7.5cm]{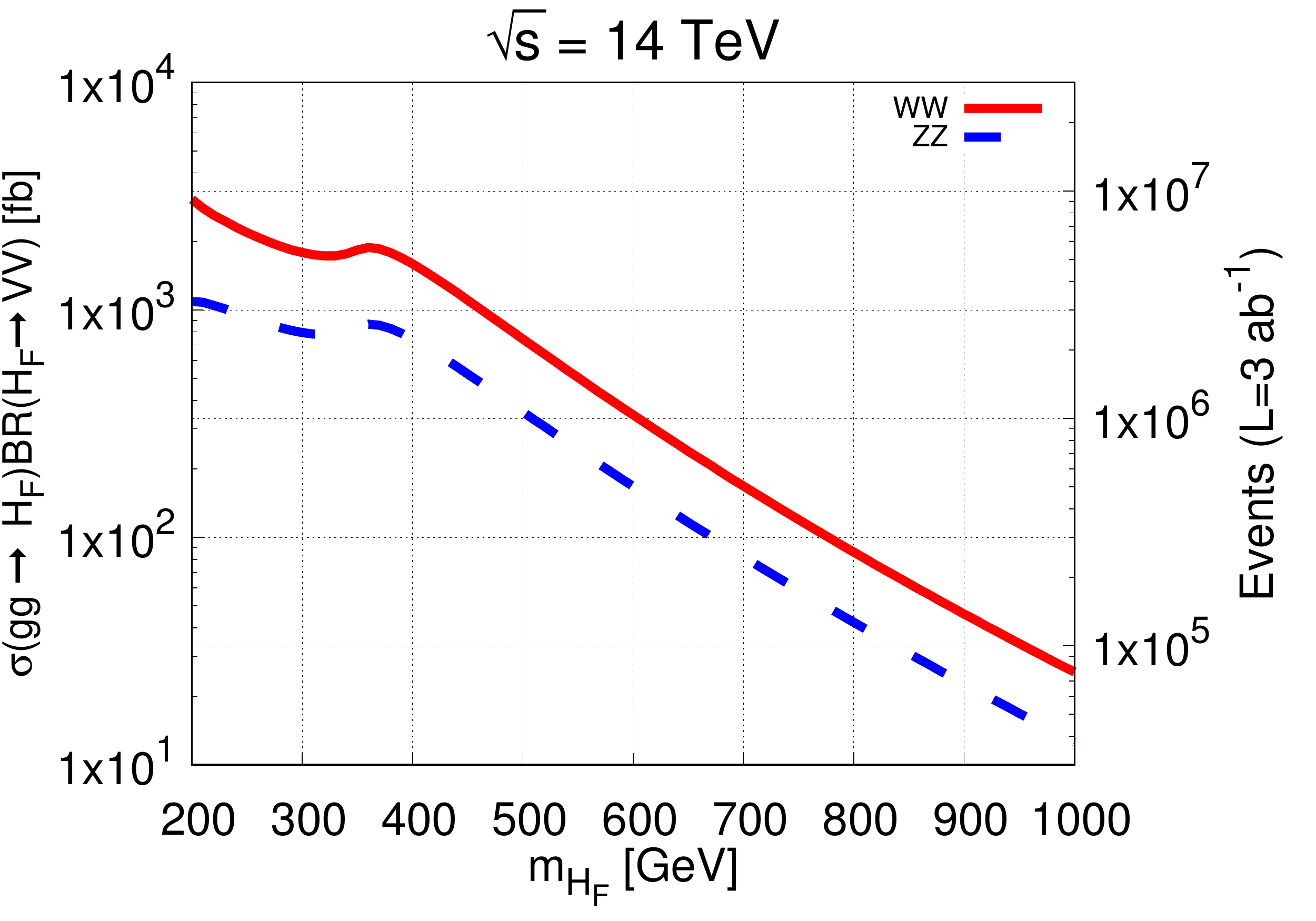}
\includegraphics[angle=270, width = 7.5cm]{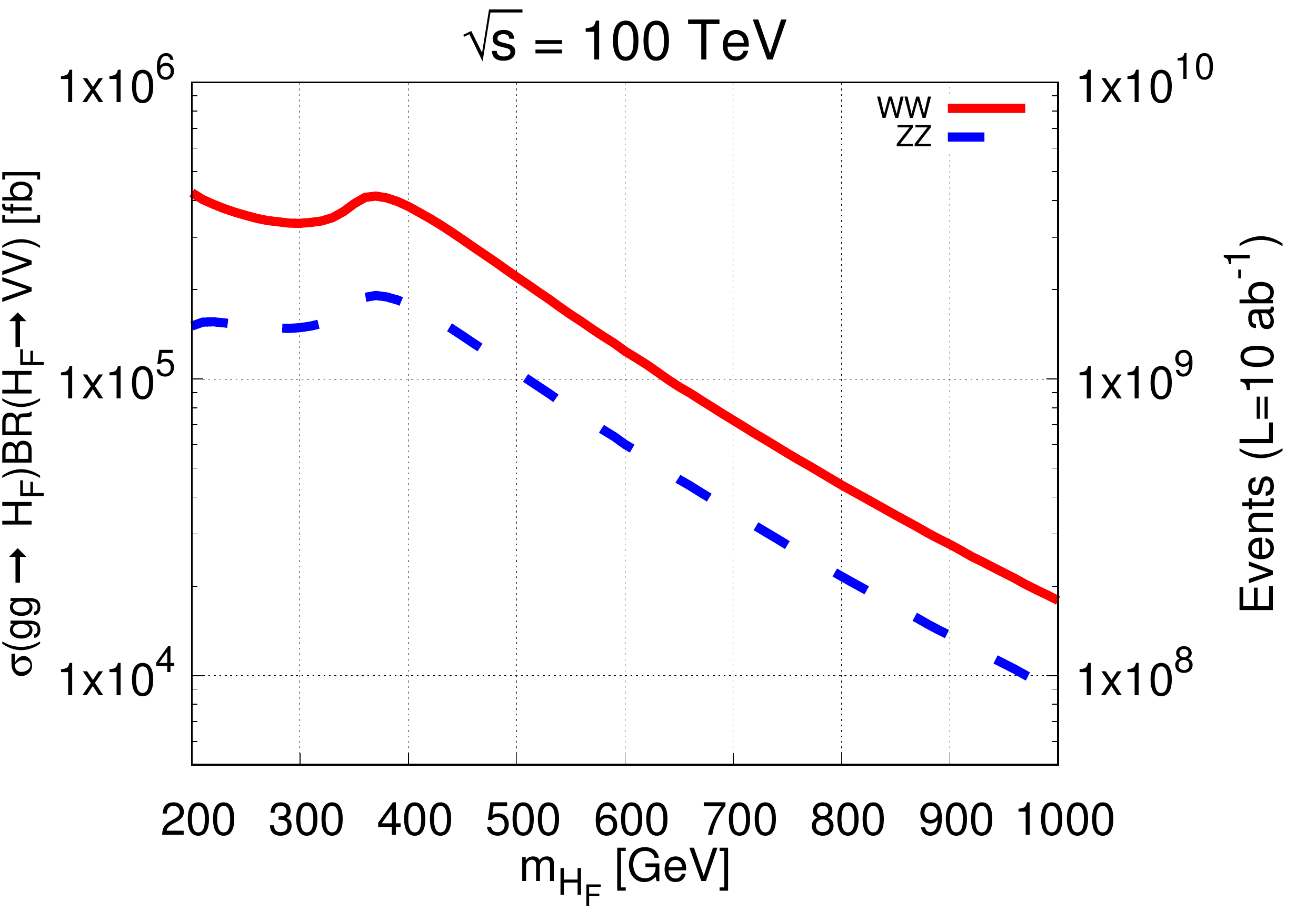}
\caption{ $pp\to H_F\to WW$ and $pp\to H_F\to ZZ$  event numbers as functions of the Flavon mass  at the LHC    (left plot) and  the FCC (right plot).}\label{WWZZChannel}
\end{figure}

Therefore, it seems troublesome that a $CP$-even Flavon with a mass greater than 300 GeV could be detected at the LHC via the $H_F\to \tau \mu$ decay channel.  In such a case, we can turn to the decay modes $H_F\to WW$ and $H_F\to ZZ$, which are the dominant ones. These decay channels seem more promising for the detection of a heavy Flavon. Along this line, to assess the potentiality of the $WW$ and $ZZ$ decay channels for the Flavon detection,  we show in Fig. \ref{WWZZChannel} the corresponding event numbers that could be produced at the LHC and a 100 TeV $pp$ collider.
We note that about $10^5$  $WW$  events would be produced at the LHC for $m_{H_F}=1$ TeV, whereas the number of $ZZ$ events would be slightly smaller. This seems more promising for the signal detection, though a more detailed analysis of the background would be required to draw a definitive conclusion.

As for the $CP$-odd Flavon, we notice that  it has a smaller $pp\to A_F\to \tau\mu$ production rate, therefore,
in order to have evidence of the $A_F\to\tau\mu$ decay, higher luminosities, of the order of $10\;\;{\rm ab}^{-1}$, would be required. This seems inaccessible for the LHC, however, we expect that the search for this decay could be possible at  the FCC.

\subsubsection{Analysis at the future 100 TeV $pp$ collider}
\begin{figure}[htb!]
\includegraphics[angle=270, width = 7.5cm]{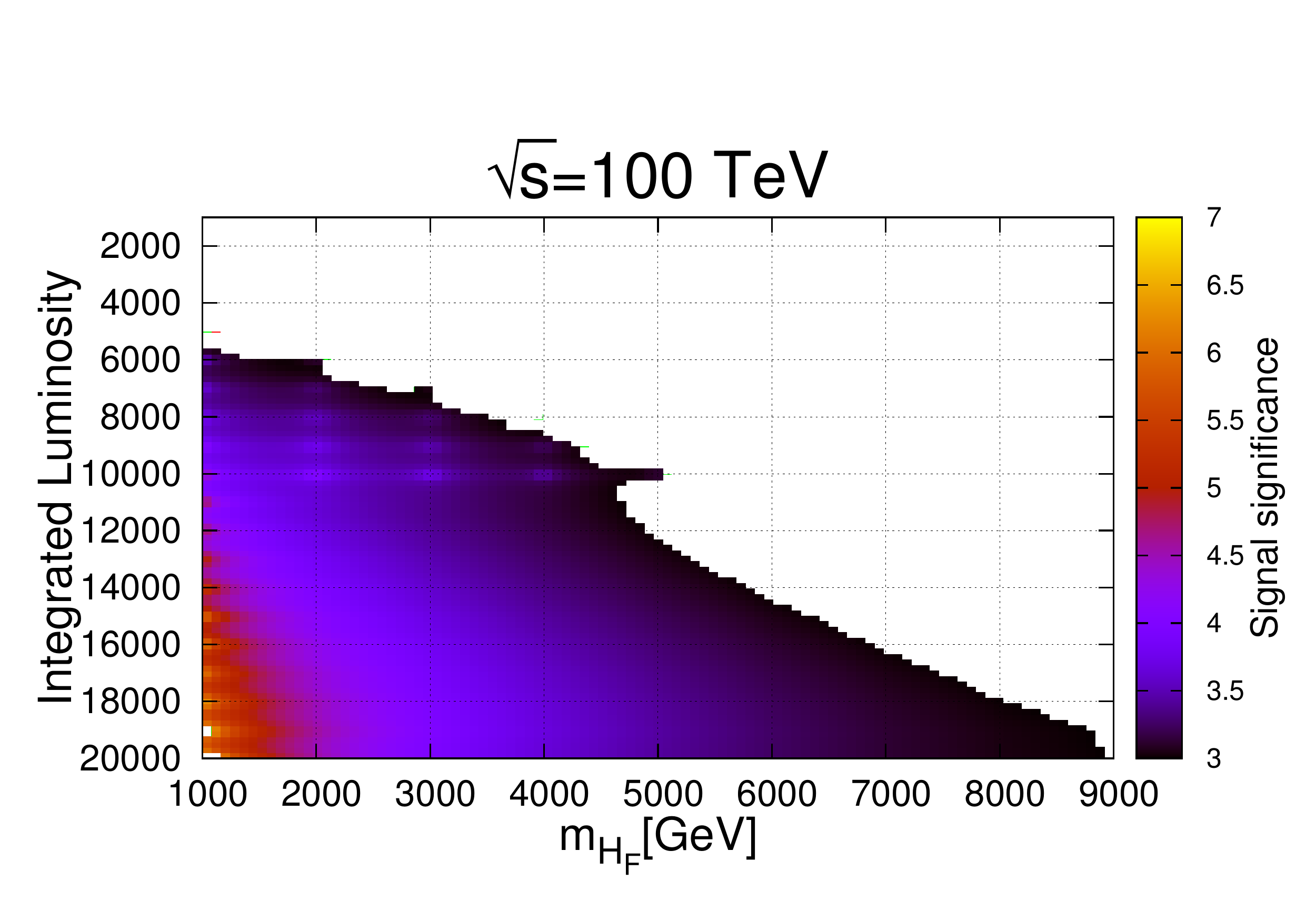}
\includegraphics[angle=270, width = 7.5cm]{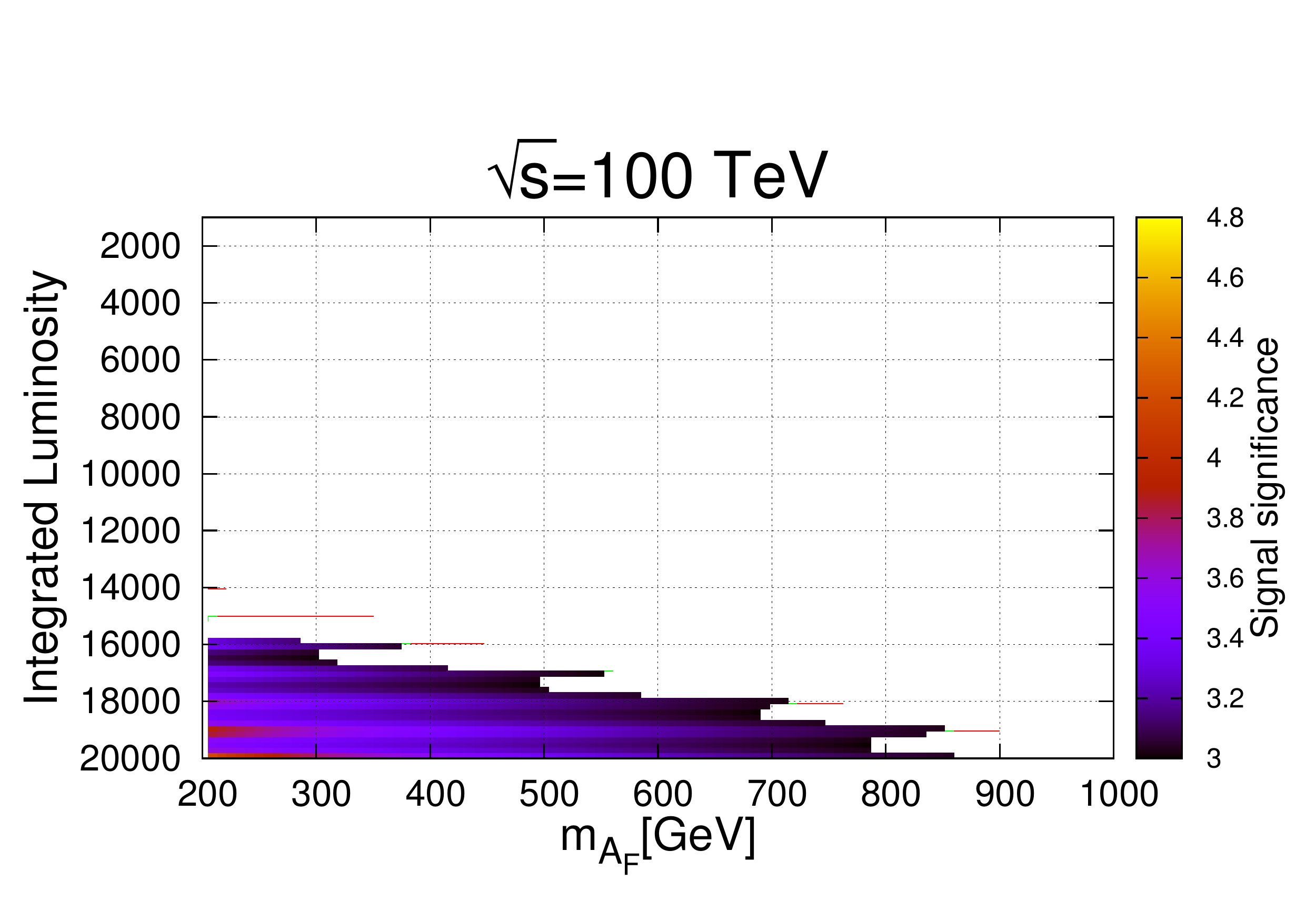}
\caption{Density plot for the signal significance as a function of the luminosity and the flavon masses $m_{H_F}$ and $m_{A_F}$.}\label{sigmaVSLumi100AF}
\end{figure}

The building of a 100 TeV $pp$ collider is under consideration \cite{Arkani-Hamed:2015vfh}. The luminosity and
center-of-mass energy are crucial factors to allow detectable LFV signatures of the FNSM. Luminosity goals for
a 100 TeV $pp$ collider are discussed by the authors of Ref. \cite{Arkani-Hamed:2015vfh}. Again, we consider the values of the Table \ref{tabla lumi}, although up to 30 ab$^{-1}$ might be reached. In our analysis we consider the conservative kinematic cuts in order to give a general overview assess how our signal could be searched for. The applied cuts to both the signal and
background are shown in Table \ref{100TeVcuts}, whereas in Fig. \ref{sigmaVSLumi100AF} we show the signal significance as a  function of the luminosity and the Flavon mass $m_{\phi}$, with $\phi=H_F,\,A_F$.
We observe that at a 100 TeV $pp$ collider with integrated luminosity of 10 ab$^{-1}$, it would be possible to probe $CP$-even Flavon  masses in the multi-TeV range, up to about 4 TeV. We also notice that $CP$-odd Flavon masses could be searched until the order  $\sim$1 TeV.

\begin{table}
\caption{Kinematic cuts applied to the $pp\to H_F\to \tau\mu$ signal and the SM main background at a 100 TeV $pp$ collider with an integrated luminosity of $\mathcal{L}=10$ ab$^{-1}$. We use  the parameter values of Table \ref{Benchmarks}  and consider a $CP$-even Flavon  with a mass $m_{H_{F}}=1000$
GeV. We also show the corresponding event numbers obtained after the kinematic cuts, and the signal significance  $S/\sqrt{S+B}$. \label{100TeVcuts}}
\centering{}%
\begin{tabular}{|c|c|c|c|c|}
\hline
{Cut number}&\bf{Cuts} & \bf{Signal (S)} & \bf{Background (B)} & $\frac{S}{\sqrt{S+B}}$\tabularnewline
\hline
\hline
&Initial (no cuts) & 90720 &   364075164  & 4.75 \tabularnewline
\hline
1& $|\eta^e|<$2.3&         80574 &          310571620 &         4.57  \tabularnewline
\hline
 2&$|\eta^{\mu}|<$2.1&          73587 &          262283714 &          4.54 \tabularnewline
\hline
 3& 150$<M_T^e$&         45334 &          117527474 &          4.18 \tabularnewline
\hline
 4& 80$<M_T^{\mu}$&          43765 &          65354838 &          5.41 \tabularnewline
\hline
 5& 100$<P_T^e$&          34252 &          57819738 &          4.5\tabularnewline
\hline
 6& 220$<P_T^{\mu}$ &          30204&          40582493 &          4.74  \tabularnewline
\hline
 7& 100$<MET<$200&         28136&          32475178 &          4.94 \tabularnewline
\hline
\end{tabular}
\end{table}

\section{Conclusions}

In this work we have explored the possibility that the LFV  $\tau\mu$ decay channel of   $CP$-even $H_F$ and $CP$-odd $A_F$ Higgs Flavons with a mass of a few hundreds of GeVs  could be at the reach of detection at the LHC and a future 100 TeV $pp$ collider, which would serve as a possible probe of low-scale flavor models.
For the theoretical framework, we have considered  the simplest  Froggatt-Nielsen  model, with an Abelian
flavor symmetry and a $CP$-conserving Higgs sector that includes a Higgs doublet  and a Froggatt-Nielsen  complex singlet.
In this model the  $CP$-even Flavon  can mix with the SM-like Higgs boson, thereby inducing tree-level LFV
interactions mediated by the latter. In this work we concentrate instead on the LFV couplings of both the  $CP$-even and $CP$-odd Higgs Flavons. After  studying  the constraints on the parameter space of the model from low-energy LFV processes, we choose a set of benchmarks and estimate the relevant decay modes  and the production cross section of the Flavons via gluon fusion at the LHC and a future 100 TeV $pp$ collider. We then consider a set of kinematic cuts for both the signal and the SM main background. It is  found that the LHC has the potential to discover the LFV decay $H_F \to \tau \mu$ for $m_{H_F}$ between 200 and 350 GeVs provided that luminosities, of the order of 1-3 ab$^{-1}$, are achieved. In such a case other decay channels would be more appropriate to search for the signal of a Flavon at the LHC.
As far as a  future 100 TeV $pp$ collider is concerned, it would be able to probe the LFV $\tau\mu$ decay channel for Flavon masses as heavy as 10 TeVs, as long as an integrated
luminosity of at least 20 ab$^{-1}$ was available, which has been deemed viable in the literature regarding the possible construction of such a collider.  Therefore, besides other physics goals,  a 100 TeV Collider might also
work as a Flavon factory.

\acknowledgments{We acknowledge support from Conacyt and SNI (M\'exico). Partial support from VIEP-BUAP is also acknowledged. }


\appendix
\section{Flavon contributions  to  LFV $h$ and $\tau$ decays and the muon anomaly}
\label{ConstraintFormulas}

In this Appendix we present the analytical expressions necessary to obtain the constraints on the LFV  Flavons couplings  shown in Fig. \ref{ParSpace}. Although these results were meant for  $CP$-even and $CP$-odd scalar bosons, they are also valid for the Flavons.

In the FNSM, the LFV decay $h\to \tau\mu$  proceeds at the tree-level.  The  decay width can be obtained from Eq. \eqref{htoff} of Appendix \ref{DecayWidthFormulas} in the $m_h\gg m_\tau\gg m_\mu$ limit. The result  is given by
\begin{equation}
\Gamma(h\to\tau\mu)=\frac{g_{h\mu\tau}m_{h}}{8\pi}.
\end{equation}
The  CMS  collaboration reported a bound on the respective branching ratio: $BR(h\to \bar{\mu}\tau) < 1.2 \times 10^{-2}$ \cite{CMS:2016qvi}.

\begin{figure}[!hbt]
  \centering
  \includegraphics[width=10cm]{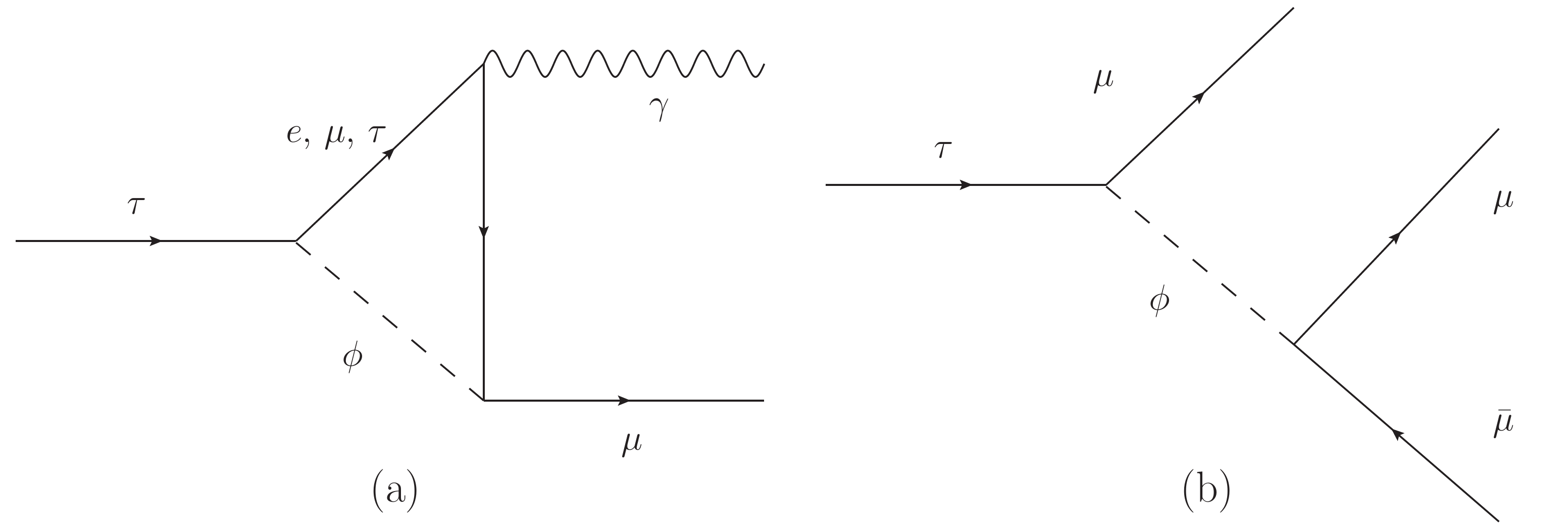}
  \caption{Feynman diagrams for the LFV decays  $\tau\to\mu\gamma$  and $\tau\to\mu\bar\mu \mu$ with exchange of a scalar boson $\phi$. We omit both the bubble diagrams for the decay  $\tau\to\mu\gamma$, which only serve to cancel the ultraviolet divergences, and the diagram where there is the exchange of the final muons in diagram (b).}\label{LFVScalarDecays}
\end{figure}

As far as the $\tau\to\mu\gamma$ decay is concerned, it arises at the one-loop level and receives contributions of the SM Higgs boson and the Flavons via the Feynman diagram of Fig. \ref{LFVScalarDecays}(a). The respective decay width is
\begin{equation}
\Gamma(\tau\to\mu\gamma)=\frac{\alpha m_{\tau}^{5}}{64\pi^{4}}\left(|C_{S}|^{2}+|C_{P}|^{2}\right),
\end{equation}
where the $C_S$ and $C_P$ coefficients stand for the contribution of $CP$-even and $CP$-odd scalar bosons, respectively, which in the limit of $g_{\phi\tau\tau}\gg g_{\phi \mu\mu} \gg g_{\phi ee}$ $(\phi=h,H_F, A_F)$ and $m_\tau\gg m_\mu\gg m_e$,
 can be approximated as \cite{Harnik:2012pb}
\begin{equation}
C_{S}  = C_P\simeq \sum_{\phi=h,H_F, A_F} \frac{g_{\phi\tau\tau}g_{\phi \mu\tau}}{12m_{\phi}^{2}}\left(3\ln\left(\frac{m_\phi^2}{m_{\tau}^2}\right)-4\right).
\end{equation}
Two-loop contributions can be relevant and the respective expressions are reported in \cite{Harnik:2012pb}. The current experimental limit on the branching ratio is $BR(\tau\to\mu\gamma)<4.4\times10^{-8}$ \cite{Olive:2016xmw}.

As for the $\tau \to \mu\bar\mu \mu$ decay, it receives contributions from the exchange of a SM-Higgs boson and the Flavons as depicted in  the Feynman diagram of Fig. \ref{LFVScalarDecays}(b). The tree-level decay width can  be approximated as

\begin{align}
\Gamma(\tau\to \mu\bar\mu \mu)&\simeq\frac{m_\tau^5}{256 \pi^3}\Bigg(\frac{S_h^2}{m_h^4}+\frac{S_{H_F}^2}{m_{H_F}^4}+\frac{S_{H_F}^2}{m_{H_F}^4}+ \frac{2 S_h S_{H_F}}{m_h^2 m_{H_F}^2}+\frac{2 S_{A_F}}{3m_{A_F}^2}\left(\frac{S_h}{m_h^2}+\frac{S_{H_F}}{m_{H_F}}\right)\Bigg),
\end{align}
where $S_\phi=g_{\phi \mu\mu} g_{\phi \mu\tau}$.
It has been pointed in Ref. \cite{Harnik:2012pb}, however, that the one-loop contribution is dominant. We refrain from presenting the corresponding expression as this process, for which the experimental limit on the respective branching ratio is $BR(\tau\to\mu\bar\mu\mu)<2.1\times10^{-8}$ \cite{Olive:2016xmw}, gives very weak constraints on the FNSM parameters.

Finally, the muon AMDM also receives contributions from the SM Higgs boson and the Flavons, which are induced by a triangle diagram similar to the diagram of Fig. \ref{LFVScalarDecays}(a) but with two external muons.  The corresponding contribution  can be approximated for $m_\phi\gg m_l$ as \cite{Harnik:2012pb}
\begin{equation}
\label{deltaamu}
\delta a_{\mu}\sim\frac{m_{\mu}}{16\pi^{2}}\sum_{\phi=h,H_F,A_F}\sum_{l=\mu,\tau}\frac{m_l g_{\phi \mu l}^2 }{m_{\phi}^{2}}\left(2\ln\left(\frac{m_{\phi}^{2}}{m_l^{2}}\right)-3\right),
\end{equation}
where  one must take into account the NP corrections to the $g_{h\mu\mu}$ coupling   only. If the Flavons are too heavy, the dominant  NP contribution would arise from the SM Higgs boson.

The discrepancy between SM theoretical prediction and the experimental value is \cite{Olive:2016xmw}
\begin{equation}
\Delta a_{\mu}=a_{\mu}^{exp}-a_{\mu}^{SM}=(2.88\pm 0.63\pm 0.49)\times 10^{-9}.
\end{equation}
Thus, the requirement that this  discrepancy is accounted for by   Eq. (\ref{deltaamu}) leads to the bound $1.32 \times 10^{-9}\le \Delta a_{\mu}\le 4.44\times 10^{-9}$ with $95\%$ C.L.
\section{Decay widths of $CP$-even and $CP$-odd scalar bosons}
\label{DecayWidthFormulas}
\subsection{$CP$-even scalar boson decays}

The most relevant decays of both  $CP$-even and  $CP$-odd scalar bosons $\phi$ have been  long studied in the literature.  We will present the relevant decay widths for the sake of completeness as they are also valid for the Flavons. We will assume that all the couplings are SM-like, other than the  $g_{H P_1 P_2}$ couplings, which stand for the couplings of a  $CP$-odd  scalar boson $H$ with the  $P_1$ and $P_2$ particles. The tree-level two-body widths are as follows.

The $\phi\to \bar{f_i} f_j$ decay width is given by
\begin{equation}
\label{htoff}
\Gamma(\phi\to \bar{f}_i f_j)=\frac{g_{\phi f_if_j}^2N_c m_\phi}{128\pi}\left(4-(\sqrt{\tau_{f_i}}+\sqrt{\tau_{f_j}})^2\right)^{\frac{3}{2}}
\sqrt{4-(\sqrt{\tau_{f_i}}-\sqrt{\tau_{f_j}})^2},
\end{equation}
where $\tau_i=4m_i^2/m_\phi^2$ and $N_c$ is the color number.  From here we easily obtain  the flavor conserving decay width.

The  decays of a heavy $CP$-even scalar boson  into pairs of real electroweak gauge bosons can also be kinematically allowed. The corresponding decay width is
\begin{equation}
\Gamma(\phi\to VV)=\frac{g_{HVV}^2 m_\phi^3}{64n_V\pi m_V^4}\sqrt{1-\tau_V}\left(1-\tau_V+\frac{3}{4}\tau_V^2\right),
\end{equation}
with $n_V=1\; (2)$ for $V=W\;(Z)$.

Other relevant  decays are those arising at the one-loop level, such as $\phi\to \gamma\gamma$ and $\phi\to gg$.  The two-photon decay width can be written as

\begin{equation}
\label{Htogammagamma}
\Gamma(\phi\to \gamma\gamma)=\frac{\alpha^2m_\phi^3}{1024\pi^3m_W^2}\left|\sum_s A_s^{\phi\gamma\gamma}\left(\tau_s\right)\right|^2,
\end{equation}
with the subscript $s$ standing for the spin of the charged particle circulating into the loop. The $A^s_{\phi\gamma\gamma}$ function is given by
\begin{equation}
A_s^{\phi\gamma\gamma}(\tau_s)=\left\{
\begin{array}{cr}
\sum_f \frac{2m_W g_{\phi ff} N_c Q_f^2}{m_f}\left[-2\tau_s\left(1+(1-\tau_s)f\left(\tau_s\right)\right)\right]&s=\frac{1}{2},\\ \\
 \frac{ g_{\phi WW}}{m_W}\left[2+3\tau_W+3\tau_W(2-\tau_W)f\left(\tau_W\right)\right]&s=1,\\ \\
 \frac{m_W g_{\phi H^-H^+}}{m_S^2}\left[\tau_{\phi^{\pm}}\left(1-\tau_{\phi^{\pm}}f\left(\tau_{\phi^{\pm}}\right)\right)\right]&s=0,
 \end{array}\right.
\end{equation}
where
\begin{equation}
f(x)=\left\{
\begin{array}{cr}
\left[\arcsin\left(\frac{1}{\sqrt{x}}\right)\right]^2&x\ge1,\\
-\frac{1}{4}\left[\log\left(\frac{1+\sqrt{1-x}}{1-\sqrt{1-x}}\right)-i\pi\right]^2&x<1.
\end{array}
\right.
\end{equation}

On the other hand, the two-gluon decay can receive contributions of quarks only and the respective decay width can be
obtained from Eq. \eqref{Htogammagamma} by summing over quarks only and making the replacements $\alpha^2\to 2\alpha^2_S$, $N_c Q_f^2\to 1$.

Formulas for other decay channels such as $\phi\to Z\gamma$, which has a considerably suppressed decay width, as well as radiative corrections for  the above decay widths can be found in the literature \cite{Djouadi:2005gi,Djouadi:2005gj}.

\subsection{$CP$-odd scalar boson decays}
The decay of a $CP$-odd scalar boson $A$ into a  pair of fermions of distinct flavor is given by
\begin{equation}
\Gamma(A\to \bar{f}_i f_j)=\frac{g_{Af_if_j}^2N_c m_A}{128\pi}\left(4-(\sqrt{\eta_{f_i}}-\sqrt{\eta_{f_j}})^2\right)^{\frac{3}{2}}\sqrt{4-(\sqrt{\eta_i}+\sqrt{\eta_j})^2},
\end{equation}
where we now use the definition $\eta_i=4m_i^2/m_A^2$. The FC decay width follows easily.

There are no decays into electroweak gauge bosons at the tree-level. On the other hand, the two-photon decay proceeds through charged fermion loops only and the corresponding decay width can be obtained from \eqref{Htogammagamma} by making the replacement $\phi\to A$, $\tau_f\to \eta_f$, and summing over fermions only, with
\begin{equation}
 A_{1/2}^{A\gamma\gamma}(\eta_f)=
\sum_f \frac{2m_W g_{A\bar{f}f} Q_f^2 N_c}{m_f}\left(-2\eta_f f(\eta_f)\right),
\end{equation}
whereas the two-gluon decay width can be obtained  by summing over quarks only and making the additional replacements $\alpha^2 \to 2\alpha^2_S$ and $N_c Q_f^2\to 1$.

\subsection{Three-body decay $H_F\to \bar{f}fh$}
\begin{figure}[!hbt]
\includegraphics[width=10cm]{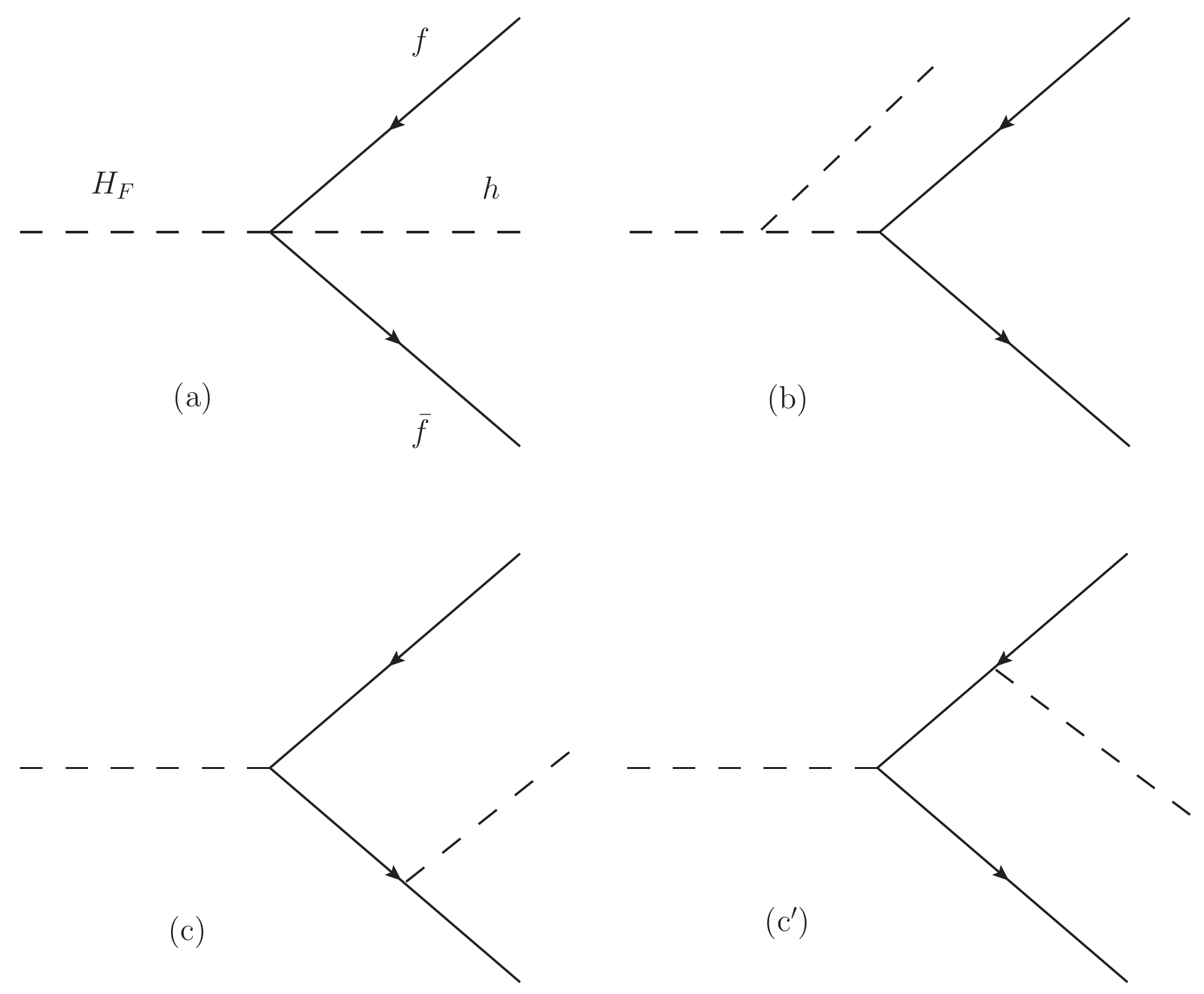}
\caption{Feynman diagrams inducing the $H\to \bar{f}fh$ decay in the FNSM.\label{Htoffh}}
\end{figure}

As far as three-body decays are concerned,  the study of the $H\to \bar{f}fh$ decay channel could be interesting as it can also have a sizeable branching ratio. This decay receives contribution from the four Feynman diagrams shown in Fig. \ref{Htoffh}. After some algebra, we can write the decay width as follows
\begin{eqnarray}
\Gamma(H_F\to h\bar{f}f)&=&\frac{m_{H_F}}{256 \pi
   ^3} \int  dx_a\int dx_b |\overline {\cal M} |^2,
\end{eqnarray}
where the integration domain is given by
\begin{equation}
2 \sqrt{x_t}\leq x_a\leq 1-x_h-2 \sqrt{x_t x_h},
\end{equation}
\begin{equation}
x_b \gtreqqless\frac{2 (1-x_h+2x_t)+x_a \left(x_a+x_h-2
   x_t-3\right)\mp\sqrt{x_a^2-4 x_t}
   \sqrt{\left(x_a+x_h-1\right)^2-4 x_h
   x_t}}{2 \left(1-x_a+x_t,
   \right)}
\end{equation}
and the average square amplitude is
\begin{align}
  |\overline {\cal M}|^2&= \frac{1}{2\left(x_a+x_b+x_h-2\right)^2}
   \left(x_a+x_b+x_h-4
   x_t-1\right)\left(
   \left(x_a+x_b+x_h-2\right)C_a+C_b\right)^2\nonumber\\&+\frac{2 }{\left(x_a-1\right)^2
   \left(x_b-1\right)^2}
   \Bigg(\left(x_a-1\right)
   \left(x_b-1\right)
   \left(x_a-x_b\right)^2-16
   \left(x_a+x_b-2\right)^2 x_t^2\nonumber\\&+4
   \left(x_a+x_b-2\right) \left(2-3
   x_b+x_a \left(4 x_b-3\right)\right)
   x_t+x_h \left(4
   \left(x_a+x_b-2\right)^2
   x_t-\left(x_a-x_b\right)^2\right)\Bigg)C_c^2\nonumber\\
&-\frac{4 \sqrt{x_t}}{\left(x_a-1\right)
   \left(x_b-1\right)}
   \left(x_a^2+2 \left(3 x_b+x_h-4
   x_t-3\right) x_a+x_b^2-4 x_h+2 x_b
   \left(x_h-4 x_t-3\right)+16
   x_t+4\right)C_a C_c\nonumber\\
&-\frac{4 \sqrt{x_t}}{\left(x_a-1\right)
   \left(x_b-1\right)
   \left(x_a+x_b+x_h-2\right)} \Bigg(x_a^2+2
   \left(3 x_b+x_h-4 x_t-3\right)
   x_a+x_b^2-4 x_h\nonumber\\&+2 x_b \left(x_h-4
   x_t-3\right)+16
   x_t+4\Bigg)C_b C_c.
\end{align}
with $x_a=(m_a/m_{H_F})^2$. Also $C_a=g_{H_F ffh}$, $C_b=g_{H_F h h}g_{h ff}/m_h^2$, and $C_c=g_{H_F ff}g_{hff}/m_h$ are the coupling constants involved in the Feynman diagrams of Fig. \ref{Htoffh}.

The general expressions for the FV decay $H_F\to f_i f_j h$ are to cumbersome unless one of the masses is neglected, so we refrain from presenting it here. The calculation was done instead via the CalcHEP software.

\subsection{The decay $A_F\to h Z$}
Although the coupling of the $Z$ gauge boson to a pair of $CP$-even and $CP$-odd scalar bosons $AhZ$  appears at the tree level in multiple Higgs doublet models, in the FNSM the scalar singlet does not couple to SM gauge bosons, so the $A_F\to h Z$ decay proceeds up to the one-loop level via fermion exchange through the Feynman diagrams of Fig. \ref{AtohZ}.
It has been pointed out recently that such a decay \cite{Bauer:2016zfj} can be relevant  to find evidences of $CP$ violation.  We have calculated the corresponding amplitude via the Passarino-Veltman reduction scheme. Using the nomenclature of Fig. \ref{AtohZ} we obtain
\begin{equation}
\label{AtohZamplitude}
\mathcal{M}(A_F \to h Z)=ig_{A_F h Z}(w_h,w_Z) (p+p_1)^\mu \epsilon(p_2)_\mu,
\end{equation}
where $w_a=m_a^2/m_{A_F}^2$, whereas  the one-loop induced coupling $g_{A_F h Z}(w_h,w_Z)$ is given as
\begin{equation}
\label{AtohZfunction}
g_{A_F h Z}(w_h,w_Z)=\frac{g}{4 \pi^2 c_W} \frac{\sum_f N_c^f g_A^f  g_{A_Fff}g_{hff} g(w_f,w_h,w_Z)}{(1-(w_h+w_Z)^2)(1-(w_h-w_Z)^2)}
,
\end{equation}
with $g_A^f=\frac{1}{2}$ ($-\frac{1}{2}$) for up (down) fermions. The $g_{A_F h Z}(w_h,w_Z)$ is given in terms of Passarino-Veltman scalar functions as follows
\begin{align}
g(w_f,w_h,w_Z)&=
2 \left(w_h w_Z+w_f
   \left(\left(w_h-1\right)^2+w_Z^2-2
   \left(w_h+1\right)
   w_Z\right)\right)C_{A_F\, h\,Z\,f}\nonumber\\&+w_f w_h
   \left(w_h-w_Z-1\right) \Delta_{h\,f\,Z}-w_f
   \left(w_h^2-\left(2 w_Z+1\right)
   w_h+\left(w_Z-1\right) w_Z\right)
   \Delta_{A_F\,f\,Z},
\end{align}
where the dimensionless functions $C_{A_F\, h\,Z\,f}$ and $\Delta_{\chi\,f\,Z}$ ($\chi=A_F\,, h$) are given by
\begin{equation}
C_{A_F\, h\,Z\,f}=m_f^2 C_0(m_{A_F}^2,m_h^2,m_Z^2,m_f^2,m_f^2,m_f^2),
\end{equation}
and
\begin{equation}
\Delta_{\chi\,f\,Z}= B_0(m_\chi^2, m_f^2, m_f^2)-B_0(m_Z^2, m_f^2, m_f^2),
\end{equation}
where $B_0$ and $C_0$ are two-point and three-point Passarino-Veltman scalar functions written in the usual notation.

The $A_F\to h Z$ decay width can be written as

\begin{equation}
\Gamma(A_F \to hZ)=\frac{|g_{A_F hZ(w_h,w_Z)}|^2m_{A_F}}{256\pi \eta_Z
}\left((4-(\sqrt{\eta_h}-\sqrt{\eta_Z})^2)(4-(\sqrt{\eta_h}+\sqrt{\eta_Z})^2)\right)^{\frac{3}{2}}.
\end{equation}
with $\eta_a=4 m_a^2/m_{A_F}^2$.
The above expressions are also valid for the decay $A_F\to H_F Z$, with the replacement $w_h\to w_{H_F}$, provided that it is kinematically allowed.

\bibliography{Revisedbiblio}

\end{document}